\documentclass[10pt]{iopart}
\usepackage{iopams}  

\usepackage{graphicx, color} 

\usepackage[colorlinks=true]{hyperref}
\hypersetup{pdfpagemode=None, linkcolor=black, citecolor=blue, urlcolor=blue}  

\usepackage[utf8]{inputenc}

\usepackage[caption=false]{subfig}
\usepackage{graphicx,bm}
\usepackage{verbatim}

\definecolor{britishracinggreen}{rgb}{0.0, 0.26, 0.15}
\definecolor{bulgarianrose}{rgb}{0.28, 0.02, 0.03}
\definecolor{darkred}{rgb}{0.90,0,0}
\definecolor{darkgreen}{rgb}{0,0.60,.2}
\definecolor{darkblue}{rgb}{0,0,1}
\definecolor{orange}{cmyk}{0,0.6,0.8,0}
\definecolor{lightblue}{rgb}{0.3,0.5,1}
\definecolor{lightgreen}{rgb}{0.4,0.80,.4}

\newcommand{\Li}{$^{6}$Li }

\begin{document}

\title[Dynamical Phase Diagram of Ultracold Josephson Junctions]
      {Dynamical Phase Diagram of 
        Ultracold Josephson Junctions
      }
\author{K. Xhani$^{1,2}$, L. Galantucci$^{1}$, C. F. Barenghi$^{1}$, G. Roati$^{2,3}$, A. Trombettoni$^{4,5}$ and N. P. Proukakis$^{1}$}
\address{
$^{1}$Joint Quantum Centre (JQC) Durham-Newcastle, School of Mathematics, Statistics and Physics, Newcastle University, Newcastle upon Tyne NE1 7RU, United Kingdom\\
$^{2}$European Laboratory for Non-Linear Spectroscopy (LENS), Universit\`{a} di Firenze, 50019 Sesto Fiorentino, Italy\\
$^{3}$Istituto Nazionale di Ottica del Consiglio Nazionale delle Ricerche (CNR-INO), 50019 Sesto Fiorentino, Italy\\
$^{4}$Department of Physics, University of Trieste, Strada Costiera
  11, I-34151 Trieste, Italy\\
$^{5}$CNR-IOM DEMOCRITOS Simulation Center and SISSA, Via Bonomea 265, I-34136
  Trieste, Italy} 
\ead{k.xhani2@ncl.ac.uk,nikolaos.proukakis@ncl.ac.uk}

\begin{abstract}
We provide a complete study of the phase diagram characterising the distinct dynamical regimes 
emerging in a three-dimensional Josephson junction in an ultracold quantum gas. Considering trapped ultracold superfluids separated into two reservoirs by a barrier of variable height and width, we analyse the population imbalance dynamics following a variable initial population mismatch. We demonstrate that as the chemical potential difference is increased, the system transitions from Josephson plasma oscillations to either a dissipative (in the limit of low and narrow barriers) or a self-trapped regime (for large and wider barriers), with a crossover between the dissipative and the self-trapping regimes which we explore and characterize for the first time. This work, which extends beyond the validity of the standard two-mode model, connects the role of the barrier width, vortex rings and associated acoustic emission with different regimes of the superfluid dynamics across the junction, establishing a framework for its experimental observation, which is found to be within current experimental reach.
\end{abstract}

\vspace{2pc}


\section{Introduction}
The Josephson effect is a direct manifestation of the macroscopic quantum phase coherence. 
First investigated in superconductors~\cite{JOS,anderson63,barone} and Helium superfluids~\cite{Sato19}, Josephson effects have also been studied for exciton-polariton condensates in
semiconductors~\cite{Lagoudakis10,Adiyatullin2017} and in dilute quantum gases, where the weak coupling between spatially separated parts can be tuned by controlling the intensity of the energy barrier between them both for ultracold bosons~\cite{JO0,JO1,Anker05,Schumm05,Shin05,Steinhauer07,JO5,LeBlanc11} and fermions~\cite{Liscience,Lidiss,Kwon84,Luick20}.
In the context of ultracold atoms, one can also realize Josephson junctions
coupling two internal states
via a weak driving field~\cite{Williams99,Smerzi03,Gross2010,Riedel2010}.
The Josephson effect is crucially used for high-precision
  measurements,
a major example being the measurement of magnetic fields
with superconducting quantum interference devices (SQUID)~\cite{Tinkham}.
The control of ultracold atomic matter has given rise to the investigation
towards analogous `atomtronic’ applications, such as the atomtronic analogue of SQUID, termed AQUID~\cite{Edwards2013,ryu_13,ryu_20}, leading to a plethora of studies
of weak link dynamics across diverse geometries and dimensionalities
(see e.g.~\cite{eckel_14,Amico_2017} and references therein).

In Josephson junctions implemented with ultracold atoms (often referred to as ultracold Josephson junctions), the Josephson current can be driven by a chemical potential difference across the junction. Unlike superconducting Josephson junctions, in their ultracold counterparts, even in the absence of any external chemical potential difference, a finite chemical potential difference can be present due to  nonlinear interactions and for a non-zero fractional population imbalance 
$z = \Delta N/N$ , where $\Delta N$ is the difference between the number of atoms in the two wells and $N$ is the total number. Such population imbalance 
is a typical parameter used to characterise ultracold Josephson junction dynamics \cite{MQST1,MQST2,MQST3}. In fact, when the initial fractional population imbalance z$_0$ is smaller than a critical value the system enters the well-known ‘plasma’ Josephson oscillations regime, featuring periodic oscillations of both relative population imbalance and relative phase about zero.
When z$_0$ instead becomes larger than a characteristic critical value, $z$ is no longer able to reach the zero value, with a bias towards the initially more populated well, such that the sign of the relative population imbalance remains unchanged in time, despite the existence of low amplitude population transfer across the weak link. This regime is known as a self-trapping regime \cite{MQST1} and it is characterised by  a ‘running’ relative phase (i.e. a relative phase which grows with time). The  occurrence of the self-trapping regime and an estimate of the critical fractional population imbalance \cite{MQST1} can be easily obtained for a Bose-Einstein condensate (BEC) in a double-well potential by writing a two-mode model starting from the mean field Gross-Pitaevskii description. Notice that in general, for long times, the self-trapped regime is eventually destroyed by thermal or quantum fluctuations \cite{ragh,Franzosi,gati2006,MQST4, MQST2,quantum_fluc,milburn1997} and/or by higher order tunneling processes \cite{Ic2}. 
Both these regimes,
discussed for ultracold bosons in double- and multi-well
potentials~\cite{MQST1,MQST2,MQST3,Trombettoni01},
have been clearly observed both in ultracold $^{87}Rb$~\cite{JO1} and  $^{39}K$~\cite{JO5}
bosonic atomic clouds trapped in a harmonic potential perturbed by a
shallow optical lattice which creates a weak link across two well-separated minima, and in the presence of a deep optical
lattice~\cite{JO0,Anker05}.

Recent experiments with fermionic $^{6}$Li have investigated
these phenomena in an elongated fermionic superfluid
across the BEC-BCS regime~\cite{Liscience,Lidiss}.
The molecular BEC regime observed in such experiments 
has a direct correspondence with the
experiments in atomic BECs. 
One of the interesting findings of 
the experiment~\cite{Liscience,Lidiss}
-- based on a thin Josephson junction --
was the explicit observation (across the entire BEC-BCS regime)
of a transition from Josephson `plasma' oscillations to a dissipative
regime with increasing initial population imbalance -- with no evidence
of the existence of self-trapping found in the probed parameter space. 
The reason for the presence of the dissipative regime, and the
  corresponding absence of the self-trapped one, can be qualitatively understood
  by observing that entering the self-trapped region from the Josephson one, the relative
  phase passes from oscillating around zero to running linearly in time, reaching
  therefore the $\pi$-value. Thus, the barrier region can be a seed
  for the creation of vortex excitations. If such excitations remain confined
  below the barrier, one may expect self-trapping to take place, while in the opposite case
  such excitations may start to propagate in the bulk of the system, giving rise to
  dissipative mechanisms. Therefore one can expect 
  that, at least for not too large values of $z_0$, there are three phases:
  Josephson plasma oscillation, self-trapping, and dissipative. 
We pause here to anticipate that one of the main goals of the present paper is to give and clarify the 
  full dynamical
  phase diagram, as a function 
  of the parameters of the system, such as
  the width of the barrier, or the anisotropy of the full three-dimensional (3D) confining potential. We will also show that, between the dissipative and the self-trapping regimes, there is an intermediate regime where the vortex rings do not 
  propagate, but there is a propagation of sound waves giving 
  rise to dissipation. Moreover, our analysis demonstrates that, for sufficiently high barriers, the dominant dissipation mechanism is not the propagation of the vortex ring {\it per se}, but instead the sound waves generated by the decaying vortex ring.

The transition from dissipationless to dissipative superflow is a
fundamental topic in its own right, whose understanding and control are
central to any potential Josephson junction applications to atomtronics. The emergence of dissipation across a Josephson junction is well-known in condensed-matter systems~\cite{caldeira}, with such dissipative process in ultracold superfluids having a close analogue to phase slips observed in superfluid
Helium~\cite{Anderson,Varoq,sato-packard-2012}.

Phase slips in ultracold Josephson junctions
have been analysed across different atomic geometries and
dimensionalities~\cite{Lidiss,Jendrzejewski,Eckel,gauthier_19,polo_19,Xhani20}.
In a three-dimensional (3D) system, our earlier work~\cite{Xhani20}
characterized the critical population imbalance for the occurrence
of such a dissipationless to dissipative transition, directly 
attributed to the phase slip associated to  the dynamical emergence of one (or more) vortex rings,
and consequent acoustic emission. Depending on the system parameters,
such vortex rings may enter the bulk condensate outside the barrier region,
with their subsequent propagating dynamics determined by an interplay
of acoustic emission, vortex-sound interactions,
kinetic energy conservation and thermal dissipation~\cite{Xhani20}. 

Vortex rings have also been discussed in the context of self-trapping:
specifically, Abad {\em et al.}~\cite{Abad} numerically related
the self-trapping regime to phase slips created by emergent vortex rings
which annihilate within the weak-link region (but outside the region of
observable condensate density). Dissipative dynamics can thus
be related to an emerging vortex ring propagating along the main axis
of the junction, and either dissipating within it, or having sufficient energy
to overcome the axial Josephson barrier and thus enter and propagate within
the bulk condensate.
For completeness, we note that this is a very distinct physical
process to the thermal-induced decay of self-trapping state observed in~\cite{Steinhauer07}
and qualitatively reproduced numerically in \cite{MQST4}. 

The above theoretical studies, combined with the existence
of several experimental studies of the Josephson effect in
ultracold superfluids observing either a transition from the
Josephson plasma oscillation regime to self-trapping~\cite{JO1,Anker05,JO5},
or a transition from  Josephson to a dissipative
regime~\cite{Liscience,Lidiss,Xhani20}, raises the interesting question of what
distinguishes between such transitions/regimes, and whether a particular
experimental set-up could be found that would allow for all three
regimes to be observed upon careful control of the relevant parameters
distinguishing between such physical regimes. 

In this work, we construct such a full phase diagram clearly demonstrating the
crossover between Josephson `plasma', dissipative and self-trapping regimes in a 3D ultracold
Josephson junction, upon careful control of the parameters (height, width)
of the barrier acting as the weak link.
Specifically, we firstly identify the parameter regime for which self-trapping
is expected to arise in an elongated harmonically-confined geometry with a
Gaussian barrier along the main trap axis
(motivated by the LENS experiment~\cite{Liscience,Lidiss} in the BEC limit),
an important feat in its own right, since only Josephson and dissipative regimes have so far been found in such a geometry.

We then generalize our studies to an isotropic harmonic trap
(i.e.~spherical condensate), and explicitly 
show -- beyond the expected  Josephson `plasma' and self-trapping regimes -- the
emergence of a dissipative regime also in such a geometry.
Our unequivocal demonstration of the existence of all three regimes
(Josephson plasma, dissipative, self-trapped) in different 3D geometries subject to careful parameter optimization paves the way for the experimental observation of such a complete phase diagram.


This paper is structured as follows: after briefly reviewing our methodology
and parameter regime (Sec.~2), we present in Sec.~3
the complete Josephson junction
dynamical phase diagram in terms of barrier height and width for
  an ultracold Josephson junction in an elongated 3D condensate.
Analysing the compressible and incompressible kinetic energy emission during
the superflow, and the properties of the vortex rings -- when emitted --
we characterize the microscopic processes controlling the regime crossover,
even in the absence of any thermal dissipation (Sec.~4). We also demonstrate
the generic nature of our 
results, by confirming their relevance in a 3D isotropic trap
(Sec.~5). Finally we discuss our findings in the context of other related
works and present our conclusions (Sec.~6). Our detailed analysis is supplemented by appropriate Appendices which provide further details into the intricate observed dynamics and crossover regions, and the relevance of the usual two-mode model.

\section{Methodology}

The 
superfluid dynamics of a 3D ultracold bosonic Josephson junction is modelled by the 
Gross-Pitaevskii equation (GPE) for the wavefunction $\psi$:
\begin{equation}\label{GPE}
i \hbar \frac{\partial \psi (\bold{r},t)}{\partial t}=-\frac{\hbar ^2}{2M} \nabla ^2 \psi (\bold{r},t)+ V_{ext}(\bold{r})\, \psi (\bold{r},t) + g |\psi(\bold{r},t)|^2 \psi (\bold{r},t)\;
\end{equation}
where $M$ is the particle mass 
and $g$ denotes the particle s-wave interaction strength.
The external trapping potential $V_{ext}(\bold{r})$ used throughout this work is based on a combination of a harmonic trap and a Gaussian barrier, leading to a double-well potential of the form:
\begin{equation}\label{Vtrap}
V_{ext}(x,y,z)=\frac{1}{2}M \left( {\omega_x}^2 x^2+ {\omega_y}^2 y^2+ {\omega_z}^2 z^2 \right)+V_0 \, e^{-2x^2/w^2}
\end{equation}
where $\omega _{x,\,y,\,z}$ are the trapping frequencies along the x, y and z directions, and the Gaussian barrier imprinted along the x-direction has a height $V_0$ and a $1/e^2$ width $w$. 

We create an initial population imbalance, $z(t=0)\equiv z_0$, between the two wells 
by adding initially a linear potential $-\epsilon x$ along the $x$ direction, and solving the GPE in imaginary time in such a tilted potential.
For simplicity we choose the initial phase difference, $\Delta \phi _0$, between the two wells to be zero.
At $t=0$, the linear potential is instantaneously removed, and the resulting population dynamics 
\begin{equation}
z(t)=\frac{N_R(t) - N_L(t)}{N}
\end{equation} 
is modelled by the time-dependent GPE, 
with
$N_L$ ($N_R$) denoting the condensate number in the left (right) reservoir,
and $N = N_L + N_R$ the total condensate number.

We consider 2 different geometries:
(i) an elongated harmonic trap (with an aspect ratio $\sim 11$), and (ii) an isotropic (spherical) harmonic trap.

The main analysis is conducted for the elongated trap, based on the parameters of Refs.~\cite{Liscience}-\cite{Lidiss}:
specifically, we use the experimental trap frequencies $\omega_x=2 \pi \times 15$ Hz, $\omega_y=2 \pi \times 187.5$ Hz, $\omega_z=2 \pi \times 148$ Hz, and a fixed particle number $N=60,000$.
The experimental atomic Josephson junction was realized by bisecting the superfluid into two weakly-coupled reservoirs by focusing onto the atomic cloud a Gaussian-shaped repulsive sheet of light of intensity $V_0$ and  a $1/e^2$ waist of $ 2.0 \pm 0.2\,\mu$m, while being homogeneous along the other 2 directions~\cite{Liscience}. 
In the BEC regime, the width of such barrier is
approximately four times the superfluid coherence length.
Here $M=2m_{\rm Li}$, where $m_{\rm Li}$ is the mass of a \Li atom, and the interaction strength $g=4 \pi \hbar^2 a _M / M$ corresponds to  an effective scattering length between the molecules $a_M \simeq 0.6 a$ (which is tunable \cite{zurn2013}), which corresponds to the molecular BEC side of the experiment with $1/(k_Fa) \simeq 4.6$, where $k_F = \sqrt{2 m E_F}/\hbar$ is the Fermi wave-vector and $a$ the interatomic scattering length.
More details on the experimental set-up can be found in Ref.~\cite{Bur14}. The validity of the GPE description on the BEC side of the BCS-BEC crossover has been discussed in 
Ref.~\cite{Lidiss},\cite{Xhani20} where it is shown that for  
our present parameters ($1/(k_F a) \simeq 4.6$) the GPE predictions agree with experimental findings.

We will also consider a spherical geometry and fix for simplicity the isotropic trap frequency  to that of the $x$-axis in the elongated experiments, i.e. $\omega _x= \omega _y=\omega _z =2 \pi \times 15$Hz, keeping the total molecule number again fixed to 60,000. 
In the elongated case, $\mu \simeq 114 \hbar \omega_x$ and the healing length $\xi = \hbar/\sqrt{\mu M} \simeq 0.067 l_x \sim 0.5 \mu$m, whereas in the spherical case $\mu \simeq 17 \hbar \omega_x$ and $\xi  \simeq 0.17 l_x \simeq 1.3 \mu$m.

\begin{figure*}[t!] 
\begin{center}
\includegraphics[width= 0.98\textwidth]{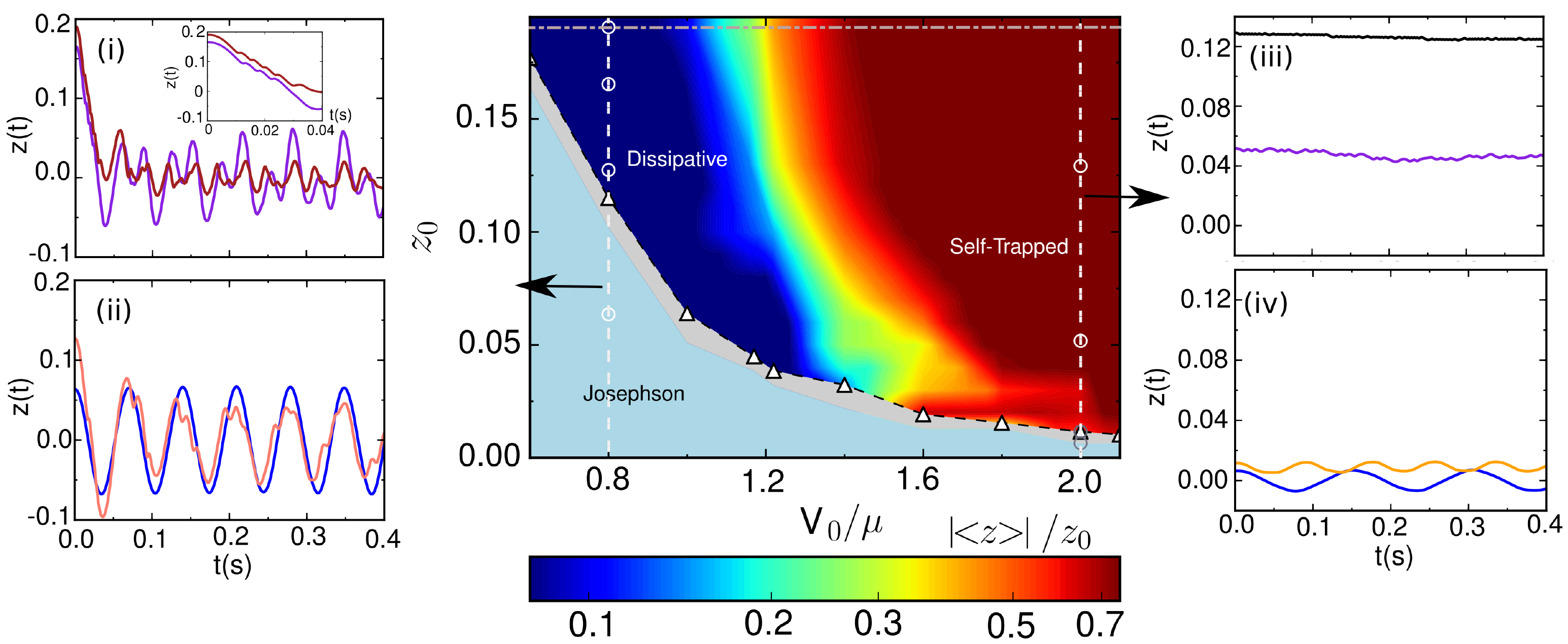}
\caption{
Main Panel: Josephson junction phase diagram of dynamical regimes arising across a Gaussian barrier of variable height $V_0/\mu\in[0.6, 2.1]$ and fixed $w/\xi = 4$ for different initial fractional population imbalances $z_0=z(t=0)$.
The Josephson plasma oscillation regime dominates for $z_0 < z_{cr}$ and, beyond a narrow transition regime (indicated by the grey shaded area) the system transitions to the dissipative ($V_0/\mu\lesssim 
1.2$), or the self-trapped ($V_0 /\mu \gtrsim 1.6$) regime, with the empty black triangles (and connected dashed black line) indicating the critical imbalance. Such regimes are characterized by the coloured regions displaying the absolute value of $\langle z(t) \rangle /z_0$ where $\langle z(t) \rangle$ is the temporal mean value of z(t) in the time interval [0, 0.4]s; the colormap is in logarithmic scale ranging between 0.08 and 0.75. In the lateral figures we plot the evolution of the population imbalance $z(t)$ for different $z_0$ and (i)-(ii) $V_0/\mu \simeq 0.8$ at $w/\xi = 4$, showing the transition from the Josephson `plasma' oscillations to the dissipative regime; and (iii)-(iv) $V_0/\mu=2$ at $w/\xi=4$, showing the transition from the Josephson to the 
self-trapping regimes.
The location of such points on the phase diagram are highlighted by the hollow white points (with the connecting white vertical lines simply a guide to the eye).
%
The horizontal dot-dashed line at the top of the phase diagram indicates the value of $z_0=0.19$ chosen for the study of the vortex ring dynamics.
}
\label{fig:w_4ksi}
\end{center}
\end{figure*}


The separation induced by the barrier, and thus the tunnelling energy across the two wells depends on two parameters: its height $V_0$ and width $w$.
Physically, it is useful to 
have them in their dimensionless ratios $V_0/\mu$ and $w/\xi$. 
The system  exhibits different behaviour across the junction 
depending on whether  
$V_0$ is much larger or smaller than the chemical potential.

In this work, we identify the different dynamical regimes across the Josephson junction by independently varying both parameters $V_0/\mu$ and $w/\xi$, thus ranging from the limit of narrow/low barriers to wide/high barriers.
Firstly, we consider the effect of changing $V_0 / \mu\in[0.6,\,2.1]$ for the fixed (thin) experimental barrier width $w/\xi = 4$ in the elongated trap. 
We then repeat our analysis in the same elongated harmonic trap for a fixed value of $V_0/\mu \sim 1.2$ and a variable barrier width in the range $w/\xi\in[4,10]$.

To verify the generality of our findings, we also consider the spherical trap geometry with the same barrier width $w/\xi = 4$, but a variable $V_0/\mu\in[0.6,\,1.8]$.
In all cases, the probed parameter space has been chosen to be broad enough, in order to reveal -- in the appropriate limits -- the emergence of all three regimes: Josephson `plasma', dissipative and self-trapped regime.

Numerically, we solve the dimensionless form of the GPE, scaling position to the harmonic oscillator length along the x direction $l_{x}=\sqrt{\hbar/M \omega_x}$, energies to the harmonic oscillator energy $\hbar \omega_x$, with densities thus scaled to $l_{x} ^{-3}$ and time in units of $1/\omega _x$.
In these units, Eq.~(\ref{GPE}) becomes:
\begin{equation}
i \frac{\partial \psi (\bold{r},t) }{\partial t}=\left(-\frac{1}{2} \nabla ^2 + V_\mathrm{ext}  + \tilde{g} |\psi(\bold{r},t)|^2 \right) \psi(\bold{r},t)  
\label{GPE_ad}
\end{equation}
where $\tilde{g}=g/(l_x ^3 \hbar \omega _x)$.
For our numerical simulations we use a numerical grid length $\left[-24,24\right]\,l_x$, $\left[-4,4\right]\,l_x$, $\left[-4,4\right]\,l_x$ along the x, y and z directions respectively, and a number of grid points of $1024\times128\times128$ for the elongated trap. For the spherical trap we use a numerical grid length $\left[-10,10\right]\,l_x$ along all three directions, and a number of grid points $256\times128\times128$, slightly biased towards the $x$ axis for better detection of the vortex rings.


\section{Dynamical regimes for the elongated trap}

We start by analyzing the complete phase diagram of emerging dynamical regimes across a Josephson junction in an elongated 3D BEC, and demonstrate clearly the dynamical behaviour across those different regimes, and also the crossover between 
them.

\subsection{Dependence on barrier height}

The LENS experiment \cite{Liscience}, conducted for a rather thin barrier $w / \xi \sim 4$ and $V_0/\mu \leq 1.2$ observed undamped Josephson plasma oscillations  and a transition to the dissipative regime.
Motivated by the unexpected absence of the self-trapping regime -- a regime 
well observed in 
other ultracold atom experiments \cite{JO1,Anker05,JO5,Zibold10} -- we extended our
numerical simulations to larger 
barrier heights.
Our previous work \cite{Xhani20} had in fact shown that the analytical two-mode model \cite{MQST1} -- on which the prediction of self-trapping is based -- does not give accurate results for $V_0/\mu \lesssim 1.2$.
(A discussion of the validity of the two-mode model for $V_0/\mu \leq 2.1$ is shown in \ref{app_1}).
Here, we extend our numerical simulations also in the direction of increasing $V_0/\mu$, and indeed find for the considered parameters the onset of the self-trapping regime for barrier height $V_0/\mu \gtrsim 1.6$. 

The full phase diagram highlighting the distinct dynamical regimes for 
a narrow barrier width $w /\xi = 4$ is shown in Fig.~\ref{fig:w_4ksi}. 
The central panel of Fig.~\ref{fig:w_4ksi} highlights the different `Josephson', `Dissipative' and `Self-Trapped' regimes, and their crossover for barrier heights in the range $[0.6, 2.1] \mu$.
While individual transitions from either Josephson plasma to dissipative, or from Josephson plasma to self-trapped had been previously 
studied both numerically and experimentally, this is the first time that all 3 regimes appear in a single phase diagram. Importantly, this is the first study displaying the gradual crossover from dissipative to self-trapped regimes.

While the Josephson plasma  regime can be clearly identified in Fig.~\ref{fig:w_4ksi} [light blue region], and despite the existence of clear regimes where dissipative, or self-trapped, dynamics dominate, the interplay between all such regimes poses challenges on how to best present the complete phase diagram.
To achieve this, the regimes beyond Josephson (i.e.~for $z_0 \geq z_{cr}$) are characterised in terms of the absolute value of $\langle z (t) \rangle/z_0$, where $\langle z (t)\rangle$ denotes the temporal mean value of the population imbalance over our entire numerically probed range (corresponding here to 0.4s). 

\begin{figure*}[t!]
\begin{center}
\includegraphics[width= 0.8\textwidth]{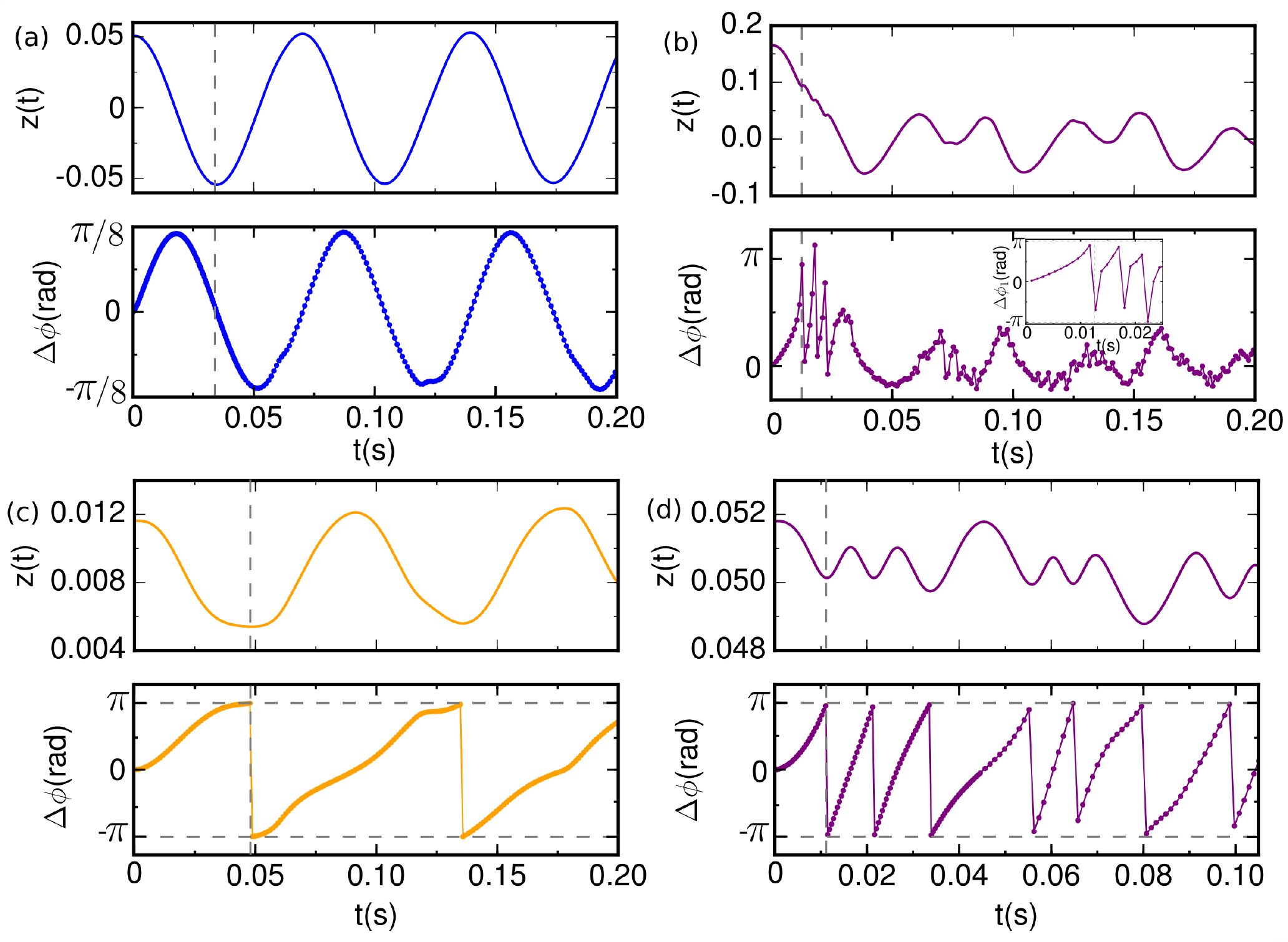}
\caption{ The temporal evolution of the imbalance and the relative phase for (a)-(b) $V_0/\mu \simeq 0.8$ and for (c)-(d) $V_0/\mu=2$ for different values of $z_0$ and at fixed $w/\xi=4$. 
The relative phase is calculated along the line $y=z=0$ near the trap center, except for the inset of (b) where the relative phase is calculated for $z=0$, but $y \simeq 0.44 l_x$. The vertical grey dashed lines show the 
time at which the population imbalance has a local minimum.} 
\label{fig:z_dphi_w_4ksi}
\end{center}
\end{figure*}

To better understand this behaviour 
, we look at the different dynamical regimes shown in Fig.~\ref{fig:w_4ksi} subplots (i)-(ii) [left panels] and (iii)-(iv) [right]. These show the dynamical evolution of the population imbalance $z(t)$ for increasing values of initial imbalance $z_0$ and for fixed $V_0/\mu \simeq 0.8$ [(i)-(ii) Josephson plasma to dissipative transition] and $V_0/\mu=2.0$ [(iii)-(iv) Josephson plasma to self-trapped transition]. In both cases, the location of the displayed cases on the phase diagram are highlighted by white hollow circles. 
The blue symmetric oscillations shown for the lowest values of $z_0$ in both (ii) and (iv) [lower panels] correspond to 
Josephson plasma oscillations with a single dominant frequency.
Note that while a dominant frequency is evident
in (iii) slightly above $z_{cr}$, 
self-trapped states well above that value in (iv) 
exhibit contributions from many frequencies. 

For $V_0/\mu \simeq 0.8$, increasing the initial population imbalance and for $z_0$ exceeding a critical value ($z_0 \geq z_{cr}$), the $z(t)$ presents `kinks' during the first transfer cycle of population across the weak link (Fig.\ref{fig:w_4ksi}(i)-(ii) and inset in Fig.\ref{fig:w_4ksi}(i)). As shown in \cite{Xhani20}, such kinks are related to the generation of vortex rings which temporarily slows down (potentially even momentarily reversing) the evolution of $z(t)$. For this low $V_0/\mu$, the vortex ring starts propagating along the long condensate axis. These kinks during the early time evolution of $z(t)$ are always observed in the dissipative regime. For $z_0=z_{cr}$ there is only one kink in the initial decay of $z(t)$ as only one vortex ring is generated [orange curve in Fig.\ref{fig:w_4ksi}(ii)] . 
The generation of the vortex ring and associated acoustic emission, combined with the subsequent sound emission from the excited decaying vortex ring eventually lead to dissipation of $z(t)$, which on average continues to oscillate about a zero mean value, but with decreasing amplitude and following a more complicated pattern, involving multiple frequencies excited during the subsequent dynamics. Increasing $z_0$ to values much higher than $z_{cr}$ [purple/red lines in Fig.~\ref{fig:w_4ksi}(i) leads to the sequential emergence of multiple vortex rings -- one at a time --, visible in the current subplots by the small-amplitude kinks during the first $\sim 40$ms. Thus, in the dissipative regime we expect $\langle z(t) \rangle \sim 0$, with such regime highlighted by the dark blue colour in Fig.~\ref{fig:w_4ksi}.
Such behaviour remains true for all values of $z_0>z_{cr}$ when $V_0/\mu \lesssim 1$, and then -- depending on the value of $z_0$ -- we see a gradual change of behaviour, the dissipative regime persisting even beyond $V_0/\mu \sim 1$ for small $z_0$.

In the opposite limit of $V_0/\mu \gg 1$, and for small enough $z_0$ which is slightly higher than $z_{cr}$, we see the clear emergence of the self-trapped regime [orange curve in Fig.~\ref{fig:w_4ksi} (iv)]. In this case, the population imbalance oscillations clearly maintain a positive value within the time interval explored, implying that despite the periodic transfer of population across the two wells, the right  well  maintains a higher population compared to the left well at long time evolution. Increasing $z_0$ to values much beyond $z_{cr}$ does maintain a dominant population in the right well, with $\langle z \rangle$ having a clear positive value, i.e.~the system exhibits self-trapping, but with additional excitations becoming relevant to the system dynamics. For rather high initial $z_0$ values compared to $z_{cr}$ (for the same $V_0/\mu$), the population imbalance changes only very mildly during such oscillations, tending towards values $\langle z(t) \rangle / z_0 \rightarrow 1^{-}$, which is labelled by the deep red regime in the phase diagram.

Due to the broad range of $z_0$ values probed, we have thus found that it is better to classify this transition in terms of the scaled population imbalance $\langle z(t) \rangle / z_0$ (rather than simply $\langle z(t) \rangle$), as already shown in the phase diagram. Moreover, the (necessarily) limited temporal evolution window probed to construct all phase diagrams in this work ($0 < t < 0.4$s) implies a weak sensitivity of calculated values $\langle z(t) \rangle$ around a zero mean in the dissipative/crossover regime: for these reasons, we consider instead the quantity $|\langle z(t) \rangle| / z_0$,
which is plotted by colour in Fig.~\ref{fig:w_4ksi} and similar subsequent figures.

The grey area between the Josephson 
oscillations and the dissipative/self-trapped regimes corresponds to a regime where the 
Josephson `plasma' oscillations with a dominant frequency are suppressed.
The intermediate region, occurring at values $V_0 / \mu > 1$ (broadly arising for our current parameters around $1.2 < V_0/\mu < 1.6$, but also dependent on $z_0$) exhibits more complicated behaviour, featuring reduced sound emission, 
a generated vortex ring which remains a `ghost' vortex (without entering the condensate), and  a complicated irregular pattern of population oscillations, during which $z(t)$ may exhibit kinks, or can change sign, but in a sufficiently irregular manner that does not meet the criteria for belonging to either of the three identified regimes.
Further details on all such crossover regions are discussed in \ref{app_2}.


To better characterise the relevant identified regimes, Fig.~\ref{fig:z_dphi_w_4ksi} shows characteristic examples of the time-evolution of the population imbalance and the corresponding relative phase for different $z_0$ and (a)-(b) $V_0/\mu \simeq 0.8$ and (c)-(d) $V_0/\mu=2$. 
The dynamics shown here correspond to a subset of plots from Fig.~\ref{fig:w_4ksi}, selected so as to better characterise both relative population and phase oscillations over the relative time period where they occur (note the different $z(t)$, $\Delta \phi$ and time axes selected to best capture the relevant behaviour).
The phase difference $\Delta \phi$ between the left and right reservoir is calculated here for $y=z=0$ and near the barrier. 
Fig.~\ref{fig:z_dphi_w_4ksi}(a) depicts the Josephson `plasma' oscillations regime where both $z(t)$ and $\Delta \phi (t) $ oscillate sinusoidally around the zero-value. 

In Fig.~\ref{fig:z_dphi_w_4ksi}(b) instead $z(t)$ decays initially in time (dissipative regime) during which $\Delta \phi (t)$ exhibits three jumps, associated with the successive nucleation of three vortex rings. However the three vortex rings do not shrink to zero at $x=0$ (i.e.~at the barrier center) but they instead propagate in the left reservoir, (i.e. $|x|\geq 2w$), and for this reason the relative phase around $x=0$ does not jump by 
$2 \pi$.
Calculating instead the relative phase at 
a non-zero value of $y$,
we observe that the relative phase (inset of Fig.~\ref{fig:z_dphi_w_4ksi}(b), bottom) jumps by almost $2 \pi$ at the time the vortex ring core is near the chosen value $y=0.44 l_x$.

Fig.~\ref{fig:z_dphi_w_4ksi}(c) depicts the established self-trapped regime typically discussed in the context of the two-mode model, and labelled as `macroscopic quantum self-trapping', or MQST: in this regime, the population imbalance exhibits regular periodic oscillations about a non-zero value, accompanied by $2 \pi$ jumps in the relative phase at the times of the z(t) local minima, with a rate captured well by the two-mode model ($\nu _\mathrm{MQST} \simeq \Delta \mu/ h \simeq 12 $Hz). 
In this work, we henceforth refer to this regime as the ``pure" self-trapped regime, in order to make a distinction to the dynamics observed for much higher initial population imbalances.

%
%
Specifically, we find that as $z_0$ increases to higher values (much beyond $z_{cr}$), not only does the amplitude of the oscillations decrease significantly, with its corresponding frequency significantly increasing, but the pattern becomes increasingly less regular: an example of this is shown in Fig.~\ref{fig:z_dphi_w_4ksi}(d) [see also \ref{app_2}]. The origin of this is the co-existence of multiple modes, corresponding to additional higher-level excitations and complicated couplings~\cite{bidasyuk_16,modugno_18}, which result in less regular dynamics than those discussed in the pure two-mode self-trapping regime \cite{MQST1}. 
Nonetheless, closer inspection even in this regime, still reveals the existence of (irregular) $2 \pi$ phase jumps at the relative population minima, thus allowing us to still characterise this regime as `self-trapped' (depicted by the dark red colour in the various phase diagrams shown in this paper).


Our analysis has clearly demonstrated, that all 3 dynamical regimes are accessible in a given (here elongated 3D) geometry, for a given barrier width, by changing the barrier height. 
Changing the barrier height naturally induces 
different behaviours 
around $V_0 /\mu \sim O(1)$, due to the effective density across the barrier. Of course, for a Gaussian barrier the junction properties depend on a combination of barrier height and width, and so our results should be reproducible when fixing geometry and barrier height, but changing width, as demonstrated below.

\subsection{Dependence on barrier width}

\begin{figure}[t!]
\begin{center}
\includegraphics[width= 0.6\columnwidth]{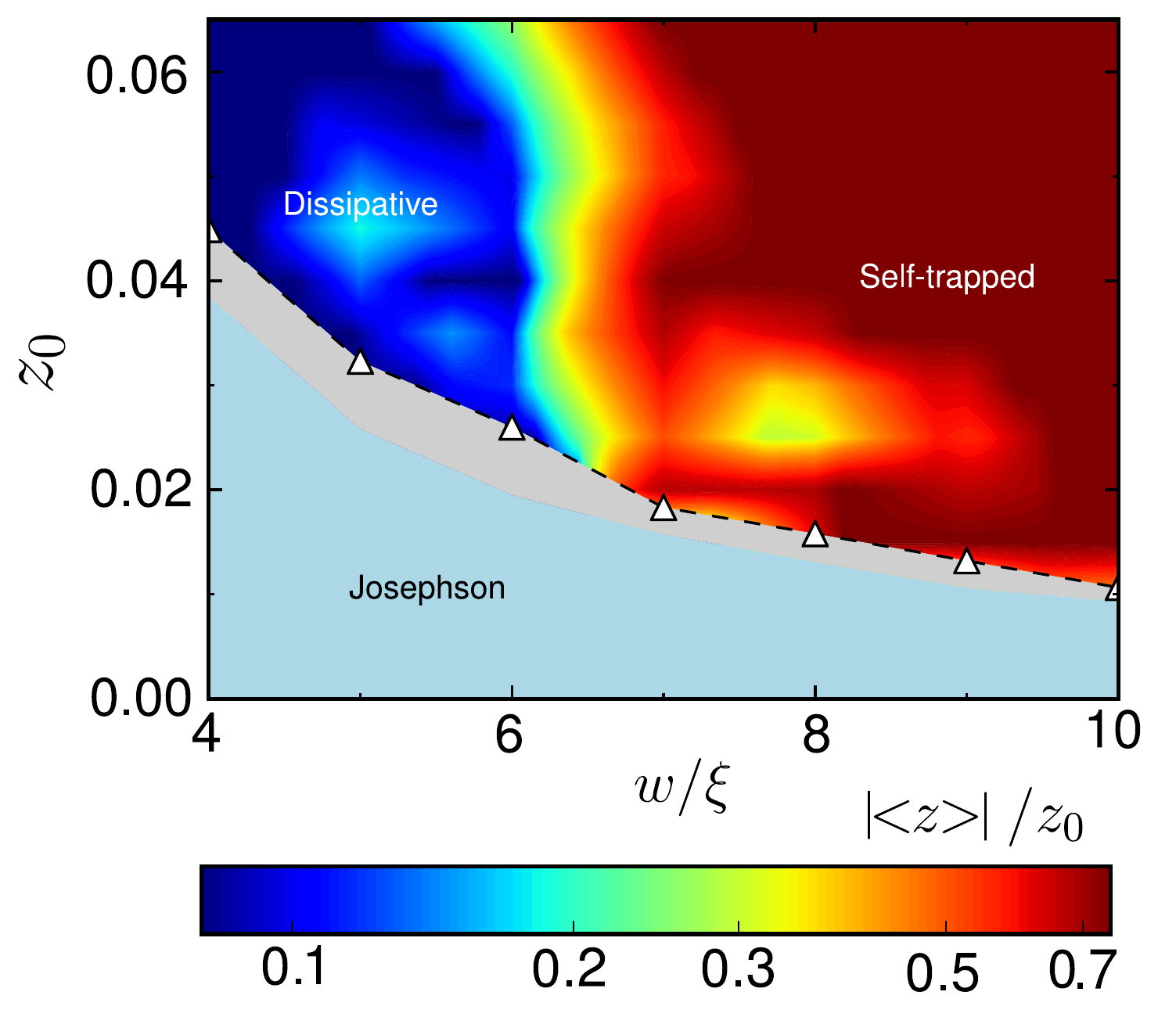}
\caption{
Josephson junction phase diagram of dynamical regimes arising across a Gaussian barrier of 
fixed $V_0/\mu=1.17$ and different values of the barrier widths in the range $[4,10] \xi$. Other symbols have the same meaning as in Fig.~\ref{fig:w_4ksi}.
} 
\label{fig:zc_vs_w_v0_1_17mu}
\end{center}
\end{figure}

Here we repeat our study of the dynamical regimes in the same elongated 3D geometry, but for fixed barrier height $V_0/\mu=1.17$ -- which we would still class as belonging to the dissipative regime 
for $w/\xi=4$ -- and different values of the barrier width compared to the healing length $\xi$.
%
The corresponding 
results are shown in Fig.~\ref{fig:zc_vs_w_v0_1_17mu}, and clearly reveal the same behaviour as found above.

When the width is narrow ($w/\xi \lesssim 6$), and for population imbalances $z_0 > z_{cr}$ such that the system transitions away from the Josephson regime, the barrier acts more like a perturbing force to the initial superfluid flow, leading to significant acoustic energy emission and vortex ring(s) generation -- with the system thus entering (in agreement with earlier findings) the dissipative regime. Increasing the barrier width at constant height, leads to a stronger effective barrier which isolates the two wells more, thus decreasing the Josephson coupling energy $E_J$ in comparison to the self-interaction energy (for fixed z$_0$) $E_C z_0 ^2 N ^2/8$: the system now transitions to the self-trapped regime. For small values $z_0$ which only slightly exceed $z_{cr}$, the system finds itself in the pure self-trapped regime, with higher $z_0$ leading to the more complicated self-trapped states discussed above.

\subsection{Critical Population Imbalance Phase Diagram}

\begin{figure}[t!]
\begin{center}
\includegraphics[width= 0.98\columnwidth]{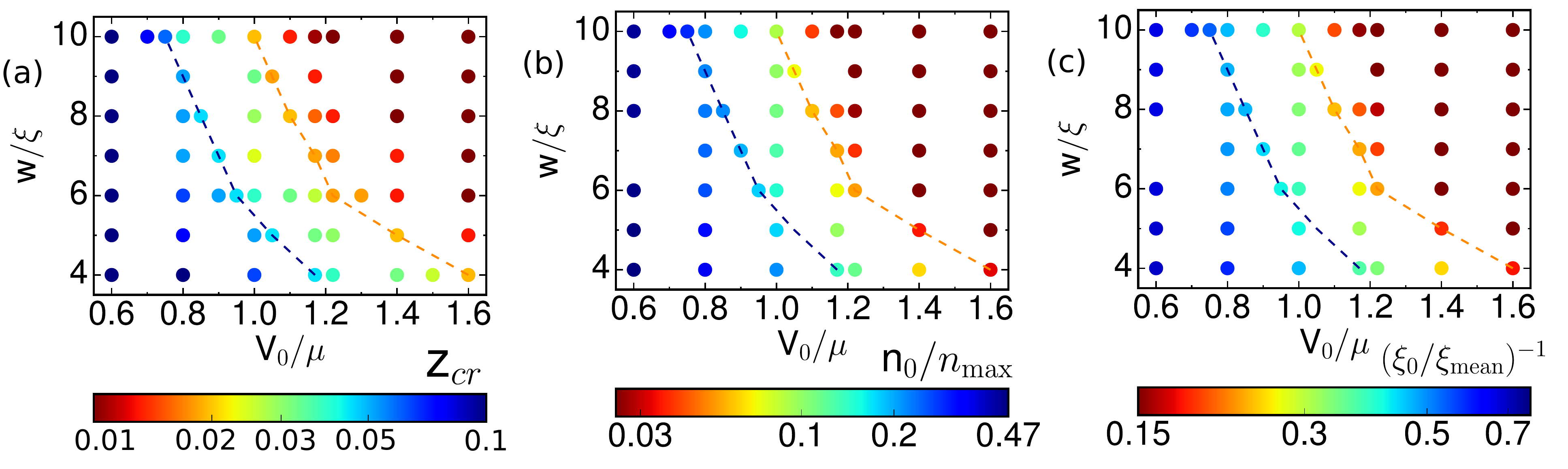}
\caption{(a) 
 Alternative phase diagram 
in the $V_0/\mu-w/\xi$ space of parameters depicting
(a) the value of the critical population imbalance, $z_{cr}$, 
(b) the scaled density, and (c) the scaled inverse coherence length, as a function of
 $V_0/\mu \in [0.6, 1.6]$ and $w/\xi \in [4, 10]$.
Their values are represented by the colorbar shown in logarithmic scale. For each value of $w/\xi$ there exists a maximum value of the barrier height $V=V_{dissip}$ (denoted by the dashed blue line) below which we find a pure dissipative regime at $z_0 \approx z_{cr}$, and another value of the barrier height $V=V_{MQST}$ (denoted by the dashed orange line), above which the system transitions to a self-trapped regime, and specifically to a pure self-trapped regime for $z_0 \approx z_{cr}$.  The numerical data between the dark-blue and the orange lines correspond to the intermediate, or `crossover', regime. In (b) the density at the trap center is scaled to the maximum value of the density in the absence of the barrier, while in (c)  the mean coherence length is correspondingly scaled to the coherence length at the trap center.
}
\label{fig:zc_all}
\end{center}
\end{figure}

To 
gain further insights on the generality 
of the results presented previously, we next investigate the dependence of the critical population imbalance, appropriately scaled densities and healing lengths 
which define the transition from Josephson to the other (dissipative; self-trapped) dynamical regimes in terms of both barrier height $V_0/\mu$ and barrier width $w/\xi$. 
This is shown in
%
Fig.~\ref{fig:zc_all}, and clearly characterizes the key parameters in terms of the transition between dissipative and self-trapped regimes.

Specifically, Fig.~\ref{fig:zc_all}(a) shows the value of the critical population imbalance, $z_{cr}$, for each probed $V_0/\mu$ and $w/\xi$ combination, which is represented by the color of the points. 
For each $w/\xi$ (and fixed $\mu$) there exists a specific (maximum) threshold value  $V_{dissip}$, such that for $V_0\leq V_{dissip}$ the system transitions from Josephson plasma to the 
dissipative regime.
Likewise, there exists a specific (minimum) threshold value of the barrier height $V_{MQST}$, such that for $V_0\geq V_{MQST}$ there is a transition to the (pure) self-trapped regime.
These crossover behaviours are respectively mapped out by the dashed blue, and dashed orange lines. 
The dissipative regime can be found for  relatively low and narrow barriers  but also for wide and small  enough  barrier heights (left of the dashed blue line).  On the other side the self-trapped regime can be achieved 
for relatively high and wide barriers but also for narrow  and large enough barrier heights (right of the dashed orange line). 
By varying the barrier heights and widths the system enters the crossover regime (for the values between the dashed blue and orange lines in Fig.~\ref{fig:zc_all}(a)), where the population imbalance features irregularly-oscillating decaying dynamics further discussed in \ref{app_2}. 

From Fig.~\ref{fig:zc_all}(a), we see that the dissipative regime occurs at a higher $z_{cr}$ than the self-trapped regime.
For the elongated experimental parameters
Refs.~\cite{Liscience,Lidiss} the critical population imbalance at which the interesting crossover dynamics emerges is rather low, placing strong constraints on its experimental observation. However, the actual value of $z_{cr}$ at such boundaries is very geometry-dependent, and we later show how an isotropic trap can significantly enhance the observable relevant region. 
%

An important comment regarding Fig.~\ref{fig:zc_all}(a) is that the value of $z_{cr}$ at which the transition to (pure) self-trapping
takes place, i.e.~along the transition line, is rather independent of the
values of $V_0/\mu$ and $w/\xi$. 
This also occurs, although to a lesser extent, along the dissipative transition line.

Interestingly, a similar qualitative picture emerges when looking at the value of the density at the trap centre, scaled to its corresponding value in the absence of the barrier. 
Another way to plot this information is as the inverse of the ratio of the coherence length calculated at the trap center to the mean value coherence length $\xi_{\rm mean}$ extracted by the maximum density in the absence of the barrier.  
These are respectively shown in Figs.~\ref{fig:zc_all}(b)-(c). 

Although these considerations clarify the weak dependence of such (dimensionless) quantities on $V_0/\mu$ and $w/\xi$, we note here that the {\em specific} value of $z_{cr}$ marking such crossovers actually depends on the trap geometry. 

Nonetheless, the above findings raise the interesting question
%
of how much the {\em structure} of the phase diagram depends upon the microscopic details of the junction and of the bulk system. We will comment more on this point in Sec.~\ref{sec_5p1}.

Before doing so,  we proceed with a detailed microscopic analysis of the observed dynamics.

\begin{figure*}[!htbp]
\begin{center}
  \includegraphics[width=.98\textwidth]{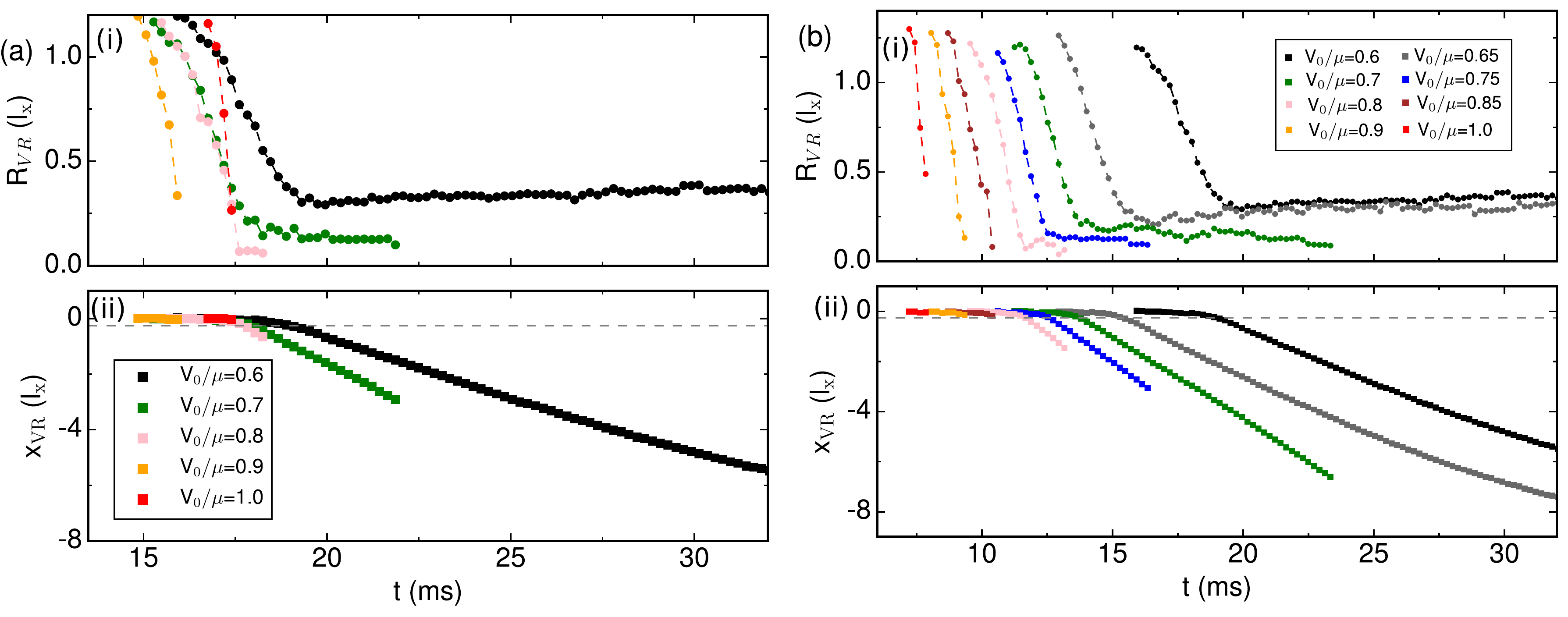}
  \caption{
  Dependence of the evolution of the first generated vortex ring for fixed value of the  barrier width $w/\xi=4$ and barrier height values $0.6 \leq V_0/\mu \leq 1$. Shown are the cases of (a) $z_0=z_{cr}$, and (b) $z_0=0.19$.
 The grey dashed horizontal lines indicate the value of the barrier width. 
 }
  \label{fig:R_z_VR_vs_V0}
\end{center}
\end{figure*}

\section{Microscopic description}

To understand the macroscopic processes described in the previous section which lead to distinct dissipative and self-trapped regimes and their crossover,
we perform here a detailed microscopic analysis of  compressible and incompressible kinetic energies and the corresponding vortex ring dynamics -- for cases where a vortex ring is indeed generated.
We apply our analysis to the entire relevant regime of $V_0/\mu \in [0.6,\,1.6]$ for the specific case of $w/\xi=4$, corresponding to the crossover shown in Fig.~\ref{fig:w_4ksi}.

For the considered value of $w/\xi$, we find that a vortex ring is generated
when $z \geq z_\mathrm{cr}$, i.e. when the initial acceleration driven by the population imbalance is such that the superfluid velocity locally exceeds a critical value within the barrier.

We start by noting that for rather
low values of $V_0/\mu$, e.g.~$V_0/\mu=0.2-0.4$, 
the large  vortex ring 
entering the condensate
%
interacts strongly, and on a relatively short timescale after commencing its axial propagation, with its image vortices \cite{galantucci-etal-2020_Leapfrogging} and, as a result of the radial anisotropy of the trap, it almost immediately breaks up into two vortex lines.
Shortly after, the vortex structure reverses its motion, travelling towards the barrier (which it can even surpass for  $V_0/\mu=0.2$).

\begin{figure*}[!htbp]
\begin{center}
 \includegraphics[width=.99\textwidth]{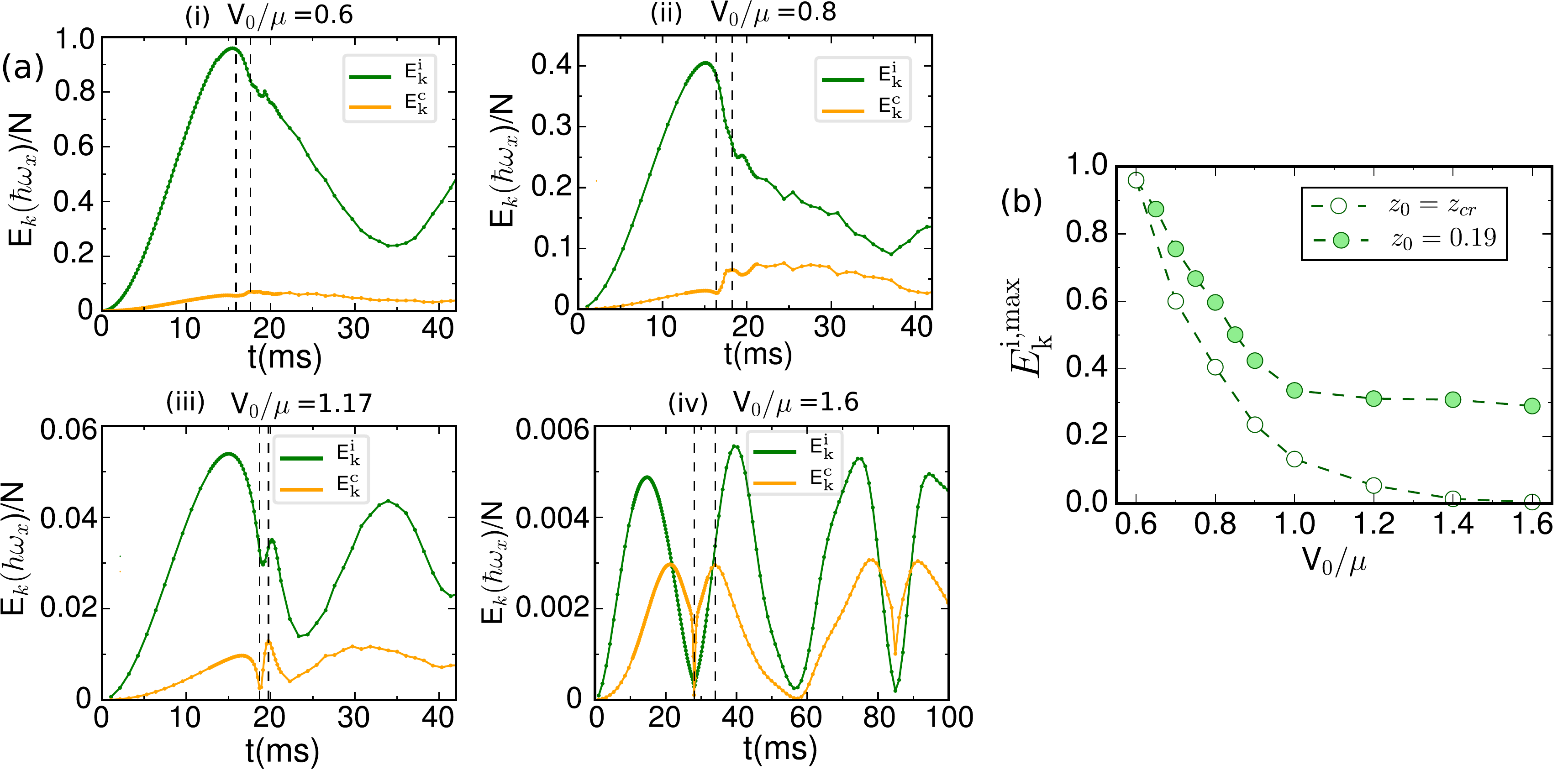}
 \caption{
 (a) The time evolution of the incompressible $E_k ^i$ (green) and compressible $E_k ^c$ (orange) kinetic energy for $z_0=z_\mathrm{cr}(V_0/\mu)$ and (i) $V_0/\mu = 0.6$, (ii) $V_0/\mu = 0.8$, (iii) $V_0/\mu = 1.17$, (iv)  $V_0/\mu = 1.6$. The vertical dashed black lines mark the boundaries of the time intervals during which $E_k ^c$ has a step-like increase, labelled here as $\Delta E_k ^c=\epsilon_c$. 
 (b) The dependence of the maximum value of  $E_k ^i$ on the barrier height $V_0/\mu$ for $z_0=z_\mathrm{cr}(V_0/\mu)$ (hollow circles) and $z_0=0.19$ (filled circles), for fixed $w/\xi=4$.
 }
  \label{fig_6_new}
  \end{center}
\end{figure*}

The 
analysis of the vortex ring motion presented in this work is therefore limited to values $V_0/\mu \ge 0.6$ for which the vortex ring does not break up immediately\footnote{Note that for $V_0/\mu=0.6$, the vortex ring still breaks into two vortex lines, but this only happens after a long evolution within the left condensate.}
In this section we characterize the dependence of vortex dynamics and relevant energies as a function of $V_0/\mu$ for two cases:
Firstly we consider
a variable initial population imbalance equal to the critical value, \textit{i.e.}~$z_0=z_{cr}(V_0/\mu)$: such a curve corresponds to the dashed black line in Fig.~\ref{fig:w_4ksi} delimiting the transition from Josephson to dissipative/self-trapped regimes.
Secondly, we consider the case of a fixed initial imbalance $z_0=0.19$, illustrated by a horizontal white dashed-dotted line at the top of Fig.~\ref{fig:w_4ksi}.
In both cases, for $V_0/\mu \lesssim 1$ a vortex is clearly observed and its dynamics within the atomic cloud is subsequently monitored 
(whereas 
for values $V_0/\mu>1$ the generated   vortex at $x=0$ always remains a `ghost' vortex outside the condensate region, where the density is negligible).

Specifically, Fig.~\ref{fig:R_z_VR_vs_V0} shows the time evolution of the radius 
$R_{\rm VR}$ (a(i), b(i)) and axial position $x_{\rm VR}$ (a(ii), b(ii)) of the clearly discernible vortex ring for $V_0/\mu  \in [0.6,\, 1.0]$ for (a) a variable $z_0=z_{cr}(V_0/\mu)$ [left], and (b) a fixed $z_0 = 0.19$ [right]. Figures 5(a) and 5(b) show a similar behaviour in this range of parameters. If $z_0=z_{cr}(V_0/\mu)$ (left),  the system 
exceeds the critical velocity once, and thus only one vortex ring is generated, while for $z_0=0.19$ multiple vortex rings can be nucleated, their number depending on the value of $z_0-z_{cr}(V_0/\mu)$. However, for simplicity, here we only study the dynamics of the first generated vortex ring.

\begin{figure*}[!htbp]
\begin{center}
 \includegraphics[width=.8\textwidth]{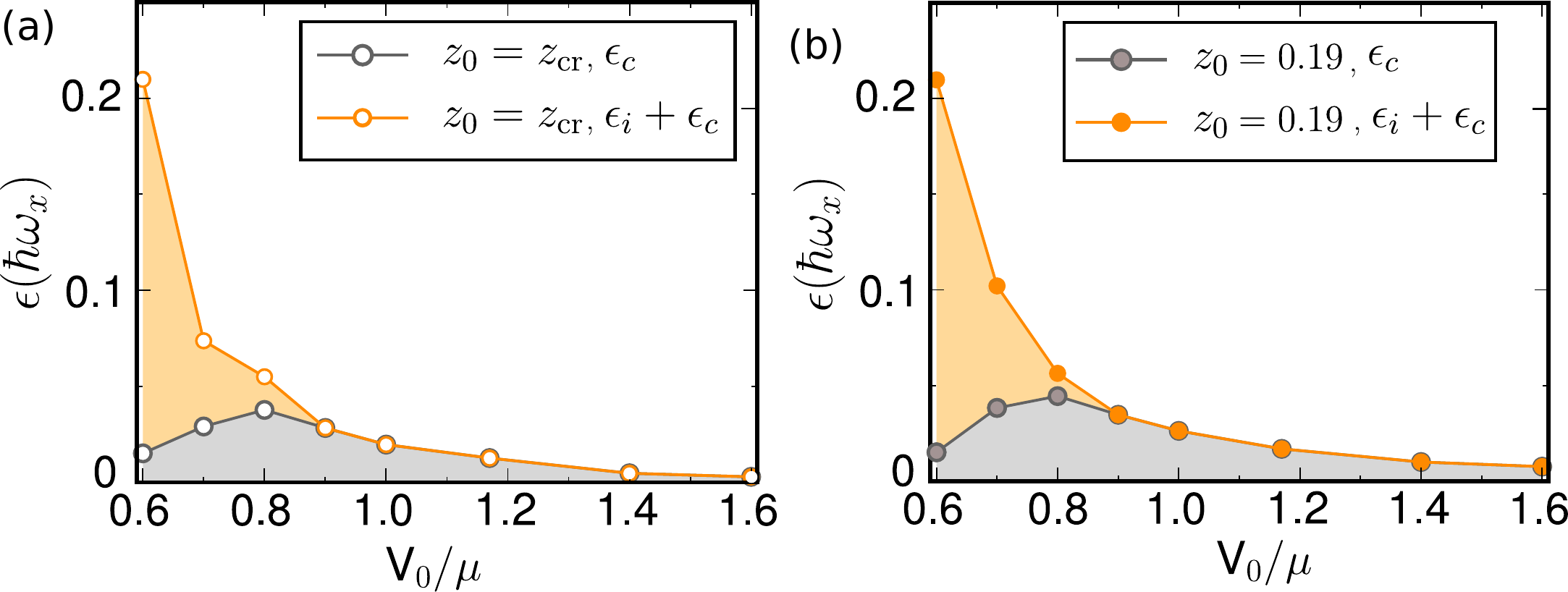}
 \caption{
 Compressible energy dissipation $\epsilon_c$ (grey circles) in the time interval bounded by the two vertical black lines in Fig.~\ref{fig_6_new}, and total energy dissipation $\epsilon_{tot}=\epsilon_c+\epsilon_i$  (orange circles) as a function of $V_0/\mu$, for fixed $w/\xi=4$ and for (a) $z_0=z_\mathrm{cr}(V_0/\mu)$ and (b) $z_0=0.19$ .
 }
  \label{fig_7_new}
  \end{center}
\end{figure*}

In the case $z_0 \simeq z_{cr}$, the vortex rings are nucleated almost after the same time interval since the start of the dynamics, independently of the value of $V_0/\mu$ (the observed minor shift is due to the numerical tracking uncertainty and numerical finite resolution in identifying z$_{cr}$). For $z_0=0.19$ and variable $V_0/\mu$ we clearly observe that larger values of $V_0/\mu$ lead to earlier nucleation of the first vortex ring; this is because the critical velocity in the barrier is reached earlier by the superfluid, due to the junction being thinner.

In both cases, we observe that as the barrier height $V_0/\mu$ decreases, the vortex ring lives longer, overcomes the barrier region and propagates further into the left well. In fact, as  $V_0/\mu$ decreases, the nucleated vortex ring has a larger energy (see below) hence a larger radius while propagating and a smaller velocity, as shown in Fig.~\ref{fig:R_z_VR_vs_V0}. For $V_0/\mu > 0.8$, the energy of the vortex ring is insufficient to overcome the barrier and thus it shrinks within the barrier itself.
This behaviour is evident in Fig.~\ref{fig:R_z_VR_vs_V0} (bottom), which shows $x_{\rm VR}$ remaining close to 0 (the motion of the vortex ring towards negative $x_{\rm VR}$ being too slow to be noticeable on this scale): for $0.8<V_0/\mu \leq 1 $ the vortex ring fails to reach $x \sim - 2 w$, the axial location in the left well at which the transversal condensate density is maximised.

Having identified the parameter regime of vortex ring generation, and characterised their dynamics, we now provide information about the energy which gives insight into the observed dynamics. Building on our earlier analysis \cite{Xhani20}, we decompose the total energy of the BEC into potential, interaction, quantum and kinetic contributions. We concentrate our attention on the kinetic energy and distinguish between the compressible $E_k ^c$ and incompressible $E_k ^i$ components, respectively defined by:
\begin{equation}
E_k ^c = \int \frac{1}{2}\left [ \left ( \sqrt{\rho} \mathbf{v} \right )^c \right ]^2 d\bold{r}
\hspace{0.5cm} {\rm and} \hspace{0.5cm}
E_k ^i = \int \frac{1}{2}\left [ \left ( \sqrt{\rho} \mathbf{v} \right )^i \right ]^2 d\bold{r}\;,
\end{equation}
where  $\vec{\nabla} \cdot (\sqrt{\rho}\vec{v}) ^i=0$ and  $\vec{\nabla}\times (\sqrt{\rho}\vec{v}) ^c=0$, with the fields $(\sqrt{\rho}\vec{v}) ^i$ and $(\sqrt{\rho}\vec{v}) ^c$  calculated via the Helmholtz decomposition \cite{nore1997, numasato,horng, griffin2019}.

We focus initially on the case $z_0=z_\mathrm{cr}(V_0/\mu)$, and calculate the time evolution of these contributions for different values of $V_0/\mu \in [0.6, \,1.6]$, with our results shown in Fig.~\ref{fig_6_new}(a).
This enables us to extract, for each $V_0/\mu$, the maximum value of  $E_k ^i$, whose dependence on $V_0/\mu$ is shown by the hollow points in Fig.~\ref{fig_6_new}(b).
For completeness, this plot also illustrates the corresponding $E_k ^i$ maxima for the fixed $z_0 = 0.19$ case (filled points).
In both cases, we clearly observe that the maximum values of  $E_k ^i$ decrease with increasing $V_0/\mu$. The energy of the vortex ring stems from the $E_k ^i$ of the flow: hence the larger $E_k ^i$, the larger the available energy for the vortex ring. This effect is visible in Fig.~\ref{fig:R_z_VR_vs_V0} (a(i), b(i)).

\begin{figure*}[!htbp]
\begin{center}
 \includegraphics[width=.99\textwidth]{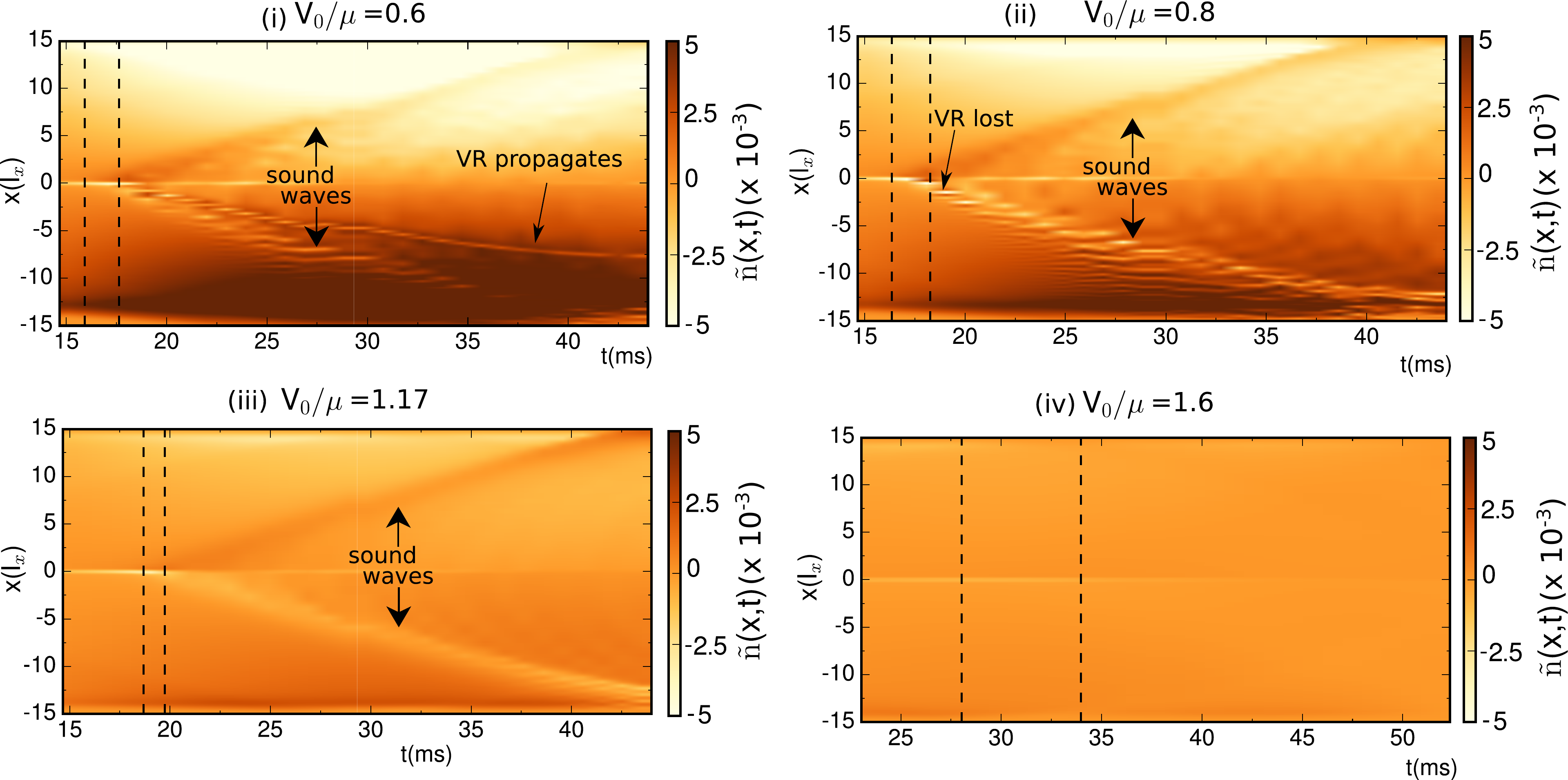}
 \caption{
 Carpet plots of renormalised density $\tilde{n}(x,t)$for $z_0=z_{cr}$ and (i) $V_0/\mu=0.6$, (ii) $V_0/\mu \simeq 0.8$, (iii) $V_0/\mu=1.17$ and (iv) $V_0/\mu=1.6$, for fixed $w/\xi=4$. Vertical dashed lines bound the time intervals between which the compressible kinetic energy is measured, and correspond to those shown in Fig.~\ref{fig_6_new}(a)(i)-(iv).
 }
  \label{fig_8_new}
  \end{center}
\end{figure*}

Until now we have been concerned with the vortex nucleation and its motion, in particular whether it can go beyond the barrier. Now we focus on the energy dissipated by the vortex. For small values of $V_0/\mu$ ($V_0/\mu\le 0.8$), the vortex overcomes the barrier and some energy of the Josephson oscillation is turned into (incompressible kinetic) energy of the vortex ring. In addition, acoustic emission takes place when the vortex, nucleated in the barrier region outside the condensate, enters the region of higher density~\cite{Xhani20}. For $V_0/\mu > 0.8$ the vortex shrinks and vanishes within the barrier, as illustrated in Fig.~\ref{fig:R_z_VR_vs_V0} (a(i), b(i)): its incompressible kinetic energy is transformed into compressible kinetic energy (sound waves).

To better characterize the compressible dissipation $\epsilon_c$, we calculate the corresponding {\em change} in the compressible kinetic energy, $\Delta E_k ^c$, experienced during the nucleation and early-stage dynamics of the vortex ring~\cite{Xhani20}: this time interval corresponds to the region between the two vertical dashed lines in panel (a)(i)-(iv). We identify this increase in compressible energy with the compressible dissipation, \textit{i.e.}  $\epsilon_c=\Delta E_k ^c$.
We observe a local maximum of $\epsilon_c$ at $V_0/\mu = 0.8$ for both $z_0=z_{cr}$ and $z_0 = 0.19$, as evident from the grey circles (and corresponding grey-shaded region) in Fig.~\ref{fig_7_new}(a)-(b). 

As for the incompressible dissipation $\epsilon_i$, we define it as the incompressible kinetic energy density integrated in a small volume surrounding the vortex ring when $x_\mathrm{VR} \simeq - l_x$~\cite{Xhani20}. We choose this value of $x_\mathrm{VR}$ as at this axial position the velocity of the Josephson flow is negligible and hence the calculated incompressible kinetic energy stems only from the vortex. When applying this definition of $\epsilon_i$ to our system, we observe that the incompressible energy of the vortex ring increases as $V_0/\mu$ decreases. This is consistent with the features illustrated in all panels of Fig.~\ref{fig:R_z_VR_vs_V0}. 
If we combine the behaviour of $\epsilon_i$ and $\epsilon_c$, we find a monotonic decrease of $\epsilon_{tot}=\epsilon_i+ \epsilon_c$ with increasing $V_0/\mu$ over the entire probed range $[0.6,\,1.6]$ for both $z_0=z_{cr}(V_0/\mu)$ [Fig.~\ref{fig_7_new}(a)] and $z_0=0.19$ [Fig.~\ref{fig_7_new}(b)]. It must be noted that for $z_0=z_{cr}$ and $V_0/\mu \simeq 0.8$ the vortex ring goes beyond the barrier but shrinks to zero and vanishes before reaching $x_\mathrm{VR} \simeq - l_x$ (where its incompressible dissipation would have been defined): its incompressible energy is totally turned into sound immediately after entering the condensate. In this circumstance, to determine $\epsilon_c$ we also consider the second step-like increase of compressible energy which is observable in Fig.\ref{fig_6_new} (a,ii) at $t\sim 20$ ms. 

In summary, our analysis demonstrates that for high barrier heights ($1 \leq V_0/\mu <1.6$), the dominant dissipation mechanism is not the propagation of the vortex ring {\em per se}, but instead the sound waves generated by the vanishing vortex ring.

To 
illustrate this effect graphically, Fig.~\ref{fig_8_new} shows `carpet plots' of the renormalised density $\tilde{n}$ along the x-direction at four different characteristic values of $V_0/\mu$ for the case $z_0=z_{cr}(V_0/\mu)$. In these plots, the density $\tilde{n}$ is evaluated by subtracting from the instantaneous density along x (for $y=z=0$), its background (equilibrium) value, \textit{i.e.} $\tilde{n}(x,t)=n_x(x,t)-n_x(x,0)$ in units of (1/$l_x ^3$). 
Subplots (i)-(iii) show clearly the propagation of sound waves in both the negative and positive x-directions for $V_0/\mu \leq 1.2$.
Subplot (i) also shows the presence of a slower moving feature, which can in fact be directly identified as
the vortex ring propagating along the negative x-axis up to -7.5 $l_x$, consistent with the vortex motion shown earlier in Fig.~\ref{fig:R_z_VR_vs_V0}(a)(ii). This vortex propagation can also be observed in Fig.~\ref{fig_8_new}(ii), but only at early times, as in this case the vortex ring vanishes rather rapidly, at $t\sim 18$ ms. The vortex ring is absent in subplots (iii), (iv) of  Fig.~\ref{fig_8_new}: specifically, in subplot (iii) we observe only the propagating sound waves while in (iv), corresponding to the self-trapped regime, the density oscillations are barely visible.

\begin{figure*}[!htbp]
\begin{center}
  \includegraphics[width=.95\textwidth]{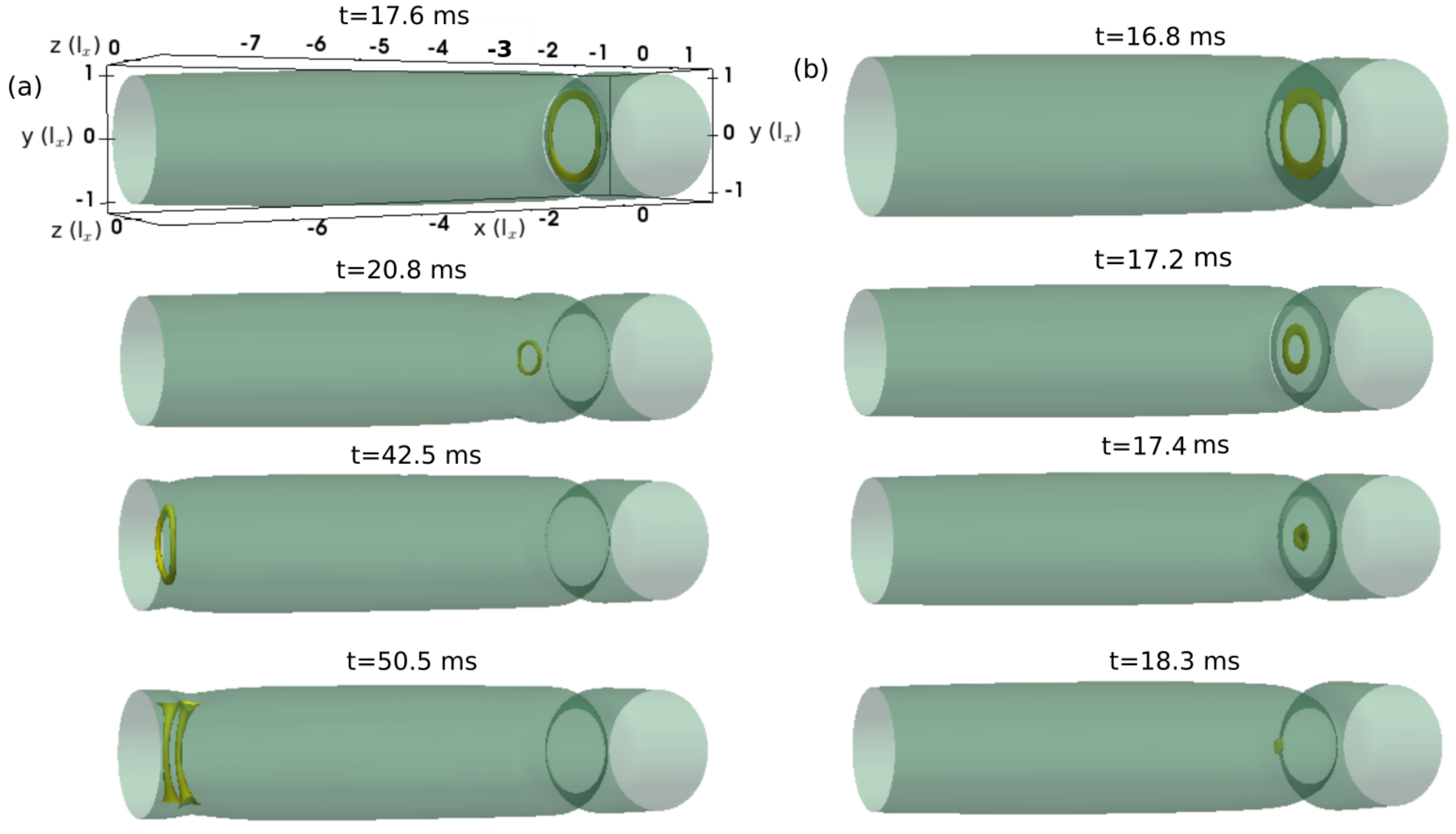}
  \caption{Snapshots of the three-dimensional isosurface density plots depicting the vortex ring generation and evolution for (a) $V_0/\mu=0.6$ and (b) $V_0/\mu \simeq 0.8$, taken at $z_0=z_{cr}(V_0/\mu)$ and $w/\xi=4$. Note the different evolution times in the two cases.}
  \label{fig:3d_elonga}
\end{center}
\end{figure*}

To clarify the vortex ring dynamics and, in particular, highlight the difference between  $V_0/\mu=0.6$ [Fig.~\ref{fig_8_new}(i)] and $V_0/\mu \simeq 0.8$ [Fig.~\ref{fig_8_new}(ii)], Fig.~\ref{fig:3d_elonga} depicts the corresponding 3D density isosurface plots.
While for $V_0/\mu=0.6$ [Fig.~\ref{fig:3d_elonga}(a)] the vortex ring propagates significantly along the negative x-axis until $x_{\rm VR} \sim - 8 l_x$, eventually breaking into two vortex lines when it reaches the boundary, for the slightly higher $V_0/\mu \simeq0.8$ the vortex ring shrinks and vanishes before reaching $x_{\rm VR} \sim -1 l_x$.

 These results complete our study of the microscopic differences between the dissipative regime for high $0.8 \le V_0/\mu \le 1.17$ and the pure self-trapped regime at $V_0/
 \mu \sim 1.6$.

We have given a detailed phase diagram for the parameter regime when different dynamical behaviours can be expected, and characterised our findings in terms of energetic considerations and vortex generation/dynamics -- in the context of an elongated 3D condensate corresponding to, and motivated by,  the LENS experimental geometry \cite{Liscience,Lidiss}. Our study would not be complete without a demonstration that our findings qualitatively hold across different experimentally-relevant geometries.

\section{Phase Diagram Extension to an Isotropic Trap \label{sec_5}}

In this section we show the broad relevance of our previously characterized phase diagram 
regimes by performing the same analysis in the context of an isotropic (spherical) trap. To make a connection with the features already studied earlier, we keep the condensate number fixed to 60,000, and all 3 harmonic trap frequencies are fixed to the previously used $\omega_x= 2 \pi \times 15$Hz. We thus set $\omega_y= \omega_z=\omega_x=2 \pi \times 15 $ Hz which (for N=60,000) gives $\mu \simeq 17 \hbar \omega _x$ and $\xi \simeq 1.3 \mu m \simeq 0.17 l_x$. 
This parameter choice -- which is within experimental reach -- has been made as it significantly increases the values of the population imbalance for which interesting dynamical crossovers can be observed by about an order of magnitude compared to the small values encountered in the elongated geometry -- thus making the observation of our findings highly experimentally relevant.

\begin{figure*}[!htbp]
\centering
  \includegraphics[width=0.6\columnwidth]{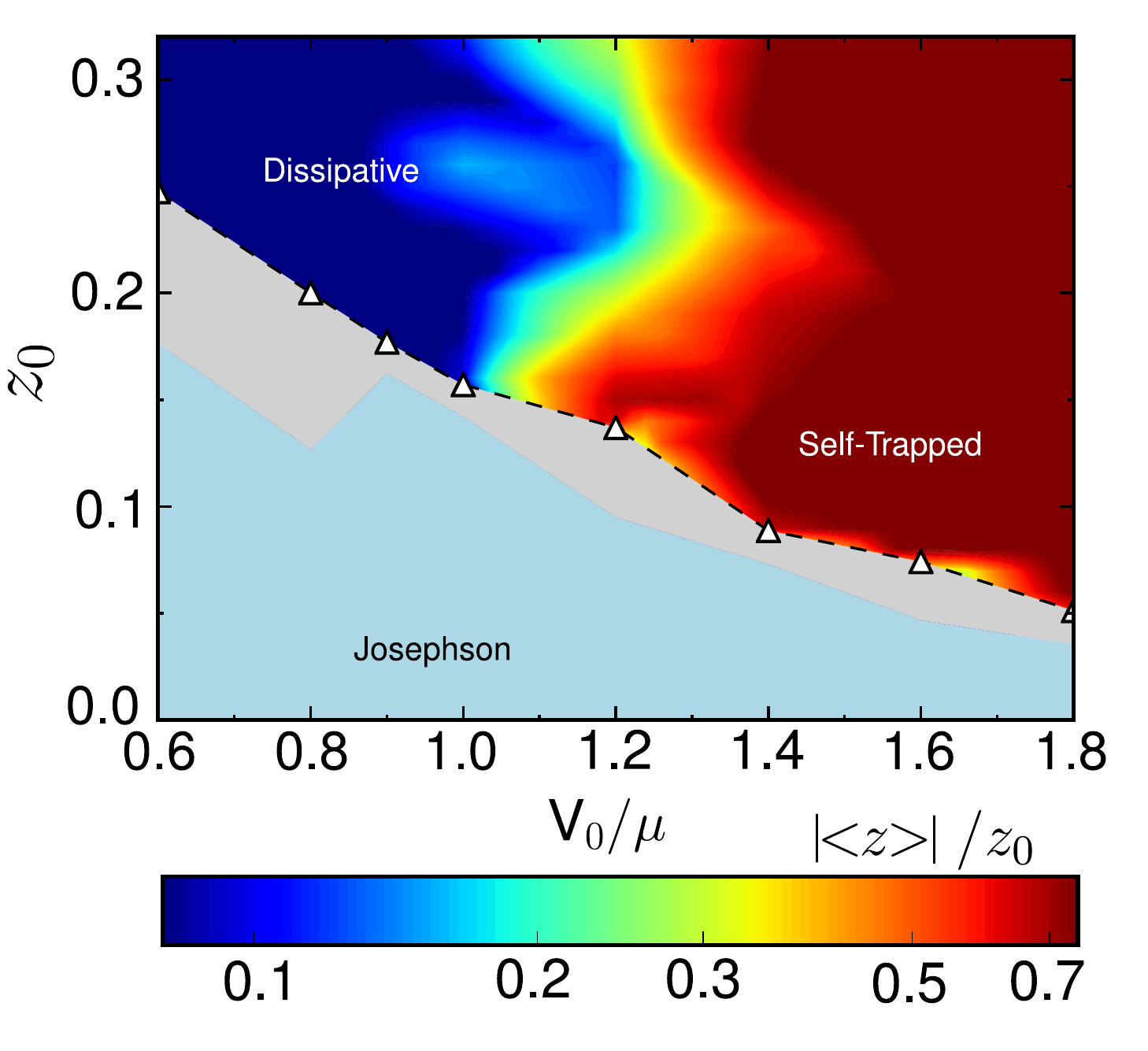}
  \caption{
  Phase diagram 
  arising across a Gaussian barrier with variable $V_0/\mu$ for
fixed $w/\xi = 4$ for an isotropic trap.
Hollow black triangles show the critical imbalance for the transition from  Josephson plasma to the dissipative regime (for $V_0/\mu \leq 1$) and to the self-trapped regime (for  $V_0/\mu \geq 1.2$).
Other symbols have the same meaning as in Fig.~1.
  }
  \label{fig:diagram_spher}
\end{figure*}

A plot revealing the emergence of the different dynamical regimes for variable $z_0$ as a function of $V_0/\mu \in [0.6,1.8]$ (similar to that of Fig.~\ref{fig:w_4ksi}.) is shown in Fig.~\ref{fig:diagram_spher}.
The important main conclusion arising from this figure is that -- despite huge differences in the values of $z_{cr}$ in relation to the elongated phase diagram  of Fig.~\ref{fig:w_4ksi}  -- qualitatively we recover the same picture.  
Values of $z_0$ below some threshold exhibit 
Josephson plasma 
sinusoidal oscillations about a zero value.
For values $z_0 \geq z_{cr}$ one instead transitions to either a dissipative regime ($V_0/\mu \lesssim 1.0$), or a self-trapped regime ($V_0/\mu \geq 1.2$), with a crossover occuring at intermediate values of $V_0/\mu$. In particular, for $0.6 \leq V_0/\mu \leq 1$, the vortex ring enters the local Thomas-Fermi surface and propagates axially into the left well, with a lifetime which decreases with increasing barrier height, as found for the elongated trap. 

The transition to the self-trapped regime is found to occur for $V_0/\mu \geq 1.2$, i.e.~at a slightly smaller value of $V_0/\mu$ with respect to the elongated trap (where it emerged around $V_0/\mu=1.6$ for the same $w/\xi=4$). Importantly, we observe that the critical imbalances for the spherical trap are higher with respect to those previously found in the elongated trap  due to the increase of the ratio of the tunneling to self-interaction energy. 

We also note in passing an interesting additional feature found within the slightly broadened grey-shaded area in this isotropic geometry.
Specifically, we have found a narrow range of intermediate values of the population imbalance $z_0$ -- located between the low values leading to single-frequency undamped Josephson plasma oscillations, and those generating the transition to the dissipative regime -- for which the observed oscillating population imbalances about a zero value can exhibit beating,
which could be attributed to enhanced coupling of the Josephson plasma oscillations to other intra-well excitations \cite{Pitaevskii}.
Indicative population imbalance plots and further details of
 the isotropic case can be found in~\ref{app_iso}.

\subsection{Discussion \label{sec_5p1}}

We comment here on the generalities of the obtained findings.
  The comparison of the results obtained for the elongated trap considered
  in Sec.~3 and the isotropic one in Sec.~5 show that the structure of the
  phase diagram is the same. Nevertheless, despite using rescaled units (e.g.,
  the barrier width $w$ in units of the healing length $\xi$ and the
  height of the potential $V_0$ in units of the chemical potential $\mu$),
  the phase diagrams are not the same, in the sense of 
  exhibiting a dependence on the
  anisotropy of the trap itself. Moreover, quantitative details also depend
  on the actual values of the trap frequencies.

  Since, from a qualitative point of view, the phase diagram structure
  depends on how vortex rings propagate, or not, in the
  bulk, one can draw an analogy (with the differences discussed below) with
  type-I/type-II superconductors \cite{Tinkham}. 
  In these latter systems, the penetration
  of an external magnetic field into the bulk of the superconducting sample
  depends on the ratio $\kappa$
  between the penetration depth and the coherence length: for
  $\kappa<\kappa_c$ (where $\kappa_c$ denotes a critical threshold value) there is a perfect
  screening of the external magnetic field, which is thus unable to enter the sample until
  the critical magnetic field is reached, while in the opposite case
  $\kappa>\kappa_c$ a partial penetration of the magnetic field
  inside the superconductor takes place through vortices~\cite{Tinkham}. Relevantly for our present discussion,
  the critical value $\kappa_c=1/\sqrt{2}$ can be considered universal,
  i.e.~independent of the microscopic details of different superconducting
  samples. Coming to our case, where the different phases depend on the
  penetration of the vortex rings in the bulk, one could be tempted
  to conclude from the type-I/type-II transitions for superconductors
  that there may exist a combination of parameters which makes the phase diagram
  independent from the system's microscopic parameters. 
  
  However, it is known that BECs 
  behave as type-II, 
  in the sense that vortices can penetrate the sample without breaking
  the superfluidity when under rotation (which is the equivalent to the
  magnetic field for neutral systems). 
  The reason for such behaviour, as
  discussed e.g.~in Ref.~\cite{Iskin11}, is that they are chargeless and the
  rotation behaves as a fictitious magnetic field, and not as a {\it real} one
  (in contrast to the magnetic field acting on superconductors which is not fictitious).

  Since BECs are of type-II, the possibility of vortex rings propagating within the bulk superfluid
  primarily depends on whether the seeded vortex inside the barrier
  can exit (overcome the barrier), or not.
  In turn, this depends sensitively on the details of the junction itself, 
  making 
 non-trivial the possibility
 to construct suitable rescaled quantities -- depending on the parameters of the system -- 
  for which the transitions between different regimes would coincide across geometrically different junctions. 
  The previous argument demonstrates the challenges in identifying appropriate dimensionless quantities, but does not
  show that one cannot in principle
  construct such suitable rescaled quantities.
  This is an important issue beyond the scope of this work, which 
  certainly deserves  further study.
  
\section{Conclusions}

We have characterised the full phase diagram describing the dynamical regimes that can emerge across a Josephson junction created by a Gaussian barrier: Josephson plasma, self-trapping, and dissipative.
Our analysis bridges the gap between numerous previous studies depicting either a transition from Josephson plasma to macroscopic quantum self-trapping, or Josephson plasma to dissipative regimes.
As expected, we have found the existence of undamped symmetric Josephson plasma oscillations 
for population imbalances below their corresponding critical values.
 Increasing the initial population imbalance across the barrier leads to a transition to a different regime, which depends on a specific combination of barrier height and width. Specifically, for relatively large barrier heights/widths, the system transitions to a self-trapped state. Once the population imbalance exceeds a critical value, it exhibits the established macroscopic quantum self-trapping regime, which features regular symmetric oscillations about a non-zero value, and a running relative phase, whereas increasing the initial population imbalance much beyond that value leads to more complicated self-trapped states with oscillations at multiple frequencies. 
In the other extreme of small barrier widths/heights, the system transitions -- with increasing $z_0$ -- to a dissipative regime, which sees the emission of acoustic (sound) energy and the generation and propagation of vortex rings, a distinctive feature associated with phase-slips known in other physical systems with Josephson junctions, and leading to the resistive  superflow. 
The critical value of $z_0$ in which dissipative behaviour is observed is always larger than the corresponding one when the system transitions (for a higher/broader barrier) to the self-trapped regime.

Our work shows that for elongated traps such as the ones studied in \cite{Liscience,Lidiss}, 
where only the transition from the Josephson plasma regime to the dissipative one was observed, the self-trapping regime can in fact also be observed for higher and wider barriers, thus making concrete predictions which can be experimentally tested. 

As a counterpart, our result indicates that for traps in which only the transition from the Josephson plasma to the self-trapped regime has been seen (such as \cite{JO1}, which had an aspect ratio $\sim 1$), the dissipative regime can also be observed by 
lowering the barrier height (to values slightly below, but still a sizable fraction of, $\mu$).

So our work suggests 
that for any geometry we can find all three dynamical regimes,
and that such regimes should be experimentally observable within a single experimental set-up by careful control of the barrier height or width.

Interestingly we also find -- beyond a smooth, and rather irregular, crossover between dissipative and self-trapped regimes -- that spherical traps have another regime that should be observable in current experiments, in which 
the coupling of the Josephson plasma frequency and other 
collective modes become relevant \cite{Pitaevskii}
and can lead to a beating. The latter becomes particularly noticeable as the system begins to transition from the pure single-frequency Josephson to the dissipative regimes.
This feature 
appears to be more pronounced in spherical geometries, rather than elongated ones. 
The spherical geometry also leads to the emergence of such features, and other crossover behaviours, at higher population imbalances, which should make such features easier to investigate experimentally.

Our results also clarify what 
distinguishes between the dissipative and macroscopic quantum self-trapping regimes regime, not just in terms of $z(t)$ and $\phi(t)$, but also in terms of vortex ring dynamics. As $V_0/\mu$ increases, the vortex rings go from a regime in which they can propagate (leaving the barrier region), to a regime where they shrink within the barrier. Whether the vortex ring leaves or not the barrier is defined by the value of incompressible kinetic energy that is present in the system. 
In the crossover between the self-trapped and dissipative regimes, the main difference comes from sound waves: specifically, in the self-trapped regime (for $z_0$ values slightly larger than  $z_{cr})$) there are practically no sound waves, while in the dissipative regime such sound waves propagate and make the condensate dissipate.

After two decades of cold atom experiments studying weak links, 
the unified description given in the present paper allows to merge different previous experimental observations together. 
Our work 
is also relevant for studying dissipation 
in a fermionic superfluid controllably tuned across the BEC-BCS crossover, which will form the basis of future work.

Finally, we observe that the results we presented are applicable to ultracold Josephson junctions for which the mean field Gross-Pitaevskii description holds. In view of experimental realizations of weak links between low-dimensional ultracold atoms, it would be very interesting to study how the dynamical phase diagram presented here is modified by the quantum fluctuations present in such systems. In the future we therefore plan to extend our study in 2D to highlight the role of dimensionality and thermal fluctuations in 2D ultracold Josephson junctions.

\ack
We thank Tilman Enss, Francesco Scazza, Augusto Smerzi and Matteo Zaccanti for useful discussions.
This work was supported by the QuantERA project NAQUAS (EPSRC EP/R043434/1), EPSRC project EP/R005192/1 and the European Research Council under GA no. 307032 QuFerm2D, the Italian MIUR under the PRIN2017 project CEnTraL.

\section*{References}
\providecommand{\newblock}{}

\clearpage
\appendix

\section{Comparison to Two-Model Model Predictions \label{app_1}}

\begin{figure*}[htb!]
\begin{center}
\includegraphics[width= 0.8\textwidth]{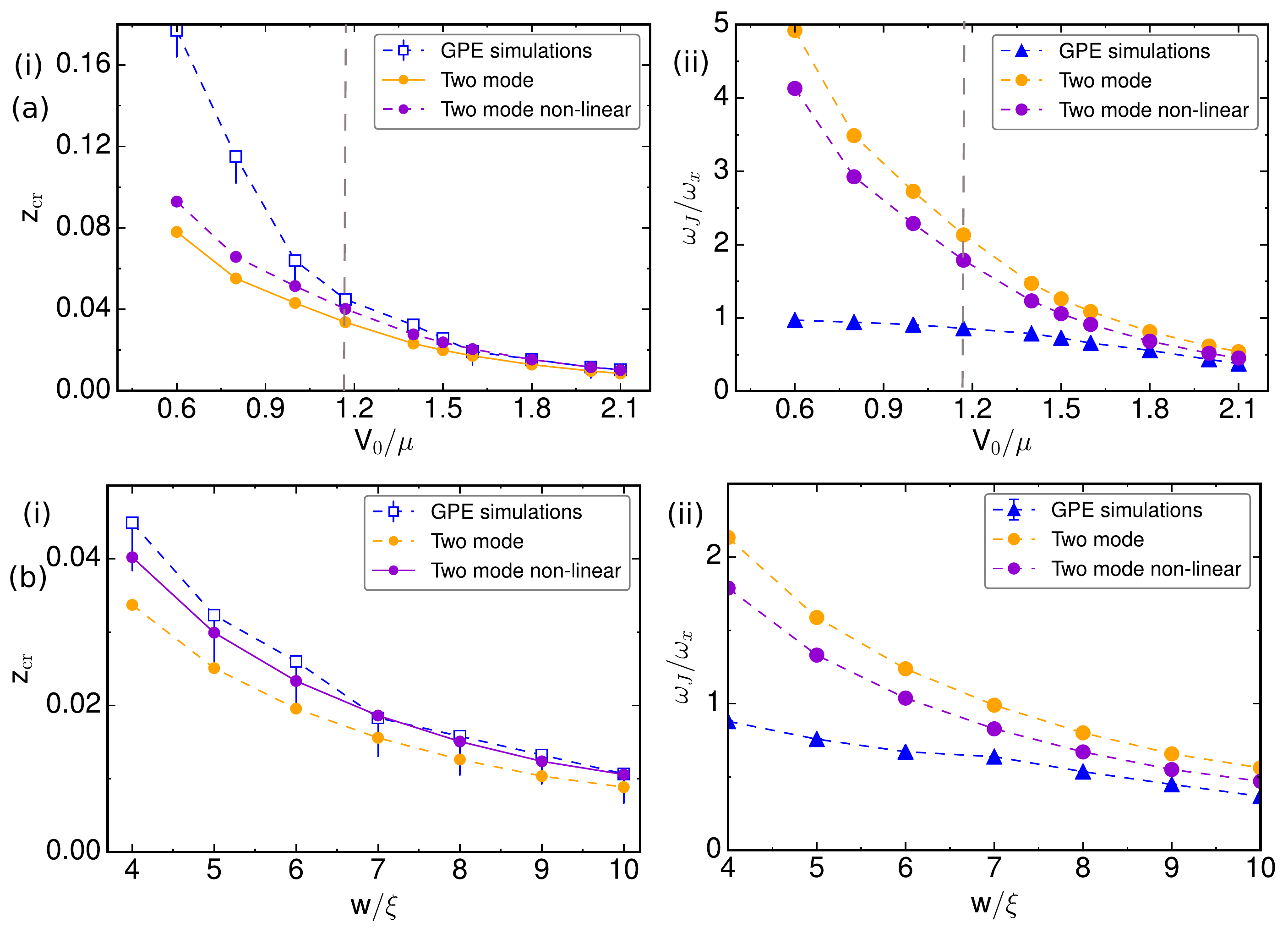}
\caption{ Critical initial population imbalance (left column, (i))
and corresponding Josephson plasma frequency (right column, (ii)) for the elongated trap, as a function of
(a) barrier height $V_0/\mu$ at fixed $w/\xi=4$, or (b) $w/\xi$ at fixed $V_0/\mu=1.17$.
All subplots show results extracted from GPE simulations (blue squares, triangles), 
and from the linear (orange circles) and nonlinear (violet circles) two-mode model.
The Josephson plasma frequencies are extracted from the GPE simulations by fitting with a sinusoidal function $z(t)$ for $z_0<z_\mathrm{cr}$. 
Vertical dashed line in (a) indicates the specific barrier height $V_0/\mu=1.17$ used in (b).
} 
\label{fig:two_mode_elongated}
\end{center}
\end{figure*}

In the standard two-mode model~\cite{MQST1,MQST3,andrea2003} the wavefunction can be expressed as  a linear superposition of the left and right condensate wave functions, i.e.
\begin{equation}
\psi(\bold{r},t) = \psi_{L} (t) \cdot \eta _{L} (\bold{r})+\psi _{R} (t) \cdot \eta _{R} (\bold{r})
\end{equation}
where $\psi_{L}(t)=\sqrt{N_{L}} e^{i \phi_{L}}$ and  $\psi_R(t)=\sqrt{N_{R}} e^{i \phi _{R}}$ and $\int \eta _i \cdot \eta _j d\bold{r}=\delta_{i,j}$, with i, j = left, right with $N_{L(R)}$ and $\phi _{L(R)}$   the number of particles and the condensate phase in the left and right well respectively. 
The left and right wavefunctions can be found from the spatially symmetric and the first antisymmetric state wavefunctions as $\eta _{R,L}= (\eta _+ \pm \eta _-)/\sqrt{2}$. The symmetric state is the ground state corresponding to zero initial imbalance and zero initial relative phase, while the antisymmetric state instead has a corresponding relative phase of $\pi$.
An atomic Josephson junction is described in terms of the on-site interaction energy $U$ and the tunneling energy $K$, from which we can extract 
the critical imbalance by using the formula $z_{cr}\simeq \sqrt{\frac{8K}{UN}}$. The tunneling energy is estimated from the difference between the  antisymmetric and the symmetric state energy  $2K \simeq E_J=(E_--E_+)=\Delta E$ and thus $z_{cr}\simeq \sqrt{\frac{4\Delta E}{UN}}$. The onsite interaction energy instead can be found from the linear two-mode model $U _\mathrm{lin}=\tilde {g} \int \eta_{L}^4 d\bold{r}$
{\em or} from the nonlinear two-mode model \cite{andrea2003} as $U_{NL}=2 (\partial \mu  /\partial N)$.

We use GPE simulations in order to find the symmetric and antisymmetric states for a linear tilted potential $\epsilon =0$, and from those we extract the tunneling energy. The extracted values of the $z_\mathrm{cr}$ from the two-mode model are shown in Fig. \ref{fig:two_mode_elongated}(a)(i) for fixed barrier width $w/\xi=4$  and barrier heights $0.6 \leq V_0/\mu \leq 2.1$ while Fig.~\ref{fig:two_mode_elongated}(b)(i) shows the corresponding results for $V_0/\mu=1.17$ and a variable barrier width $4 \leq w/\xi \leq 10$. We note that the critical imbalance from the two-mode model sets the transition to the self-trapped regime which in our case happens at fixed $w/\xi=4$ and $V_0/\mu\geq 1.6$, and  for $w/\xi \geq 7$ for fixed $V_0/\mu=1.17$. 
We also show in these subplots the corresponding extracted values of $z_{cr}$ from the GPE simulations finding good agreement for $V_0/\mu \geq 1.2$ and for all the explored barrier widths in the case of fixed $V_0/\mu=1.17$. 

The GPE prediction of z$_\mathrm{cr}$ is found by  solving again the GPE numerically, but this time with an initial shift $\epsilon$ different from zero, and thus $z_0\neq 0$. The critical imbalance is defined then by looking at the time evolution of $z(t)$. 
In the regime of the Josephson plasma oscillation and for $E_J \ll E_c$ the two-mode model prediction for the oscillation frequency is:
\begin{equation}
\omega_J\simeq\frac{\sqrt{E_J E_c}}{\hbar}=\frac{\sqrt{\Delta E U_\mathrm{lin (NL)} N }}{\hbar} \;.
\end{equation}
whose behaviour is shown in the right subplots of Fig. \ref{fig:two_mode_elongated} ((a)(ii), (b)(ii)).
Specifically, we plot $\omega_J/\omega_x$ as a function of 
$V_0/\mu$ at fixed $w/\xi=4$ [(a)(ii)] and as a function of $w/\xi$ at fixed $V_0/\mu$ [Fig.~\ref{fig:two_mode_elongated}(b)(ii)], 
showing both two-mode model predictions, and corresponding numerical GPE results.
together with the extracted values from the GPE simulations. The two-mode model predicts the Josephson plasma frequency well for $V_0/\mu \geq 1.6$ in the case of fixed $w/\xi=4$ and for $w/\xi \geq 7$ for fixed $V_0/\mu=1.17$.

\section{Further Characterization of Dynamical Regimes Crossover \label{app_2} 
}

\begin{figure*}[h!]
\begin{center}
\includegraphics[width=\textwidth]{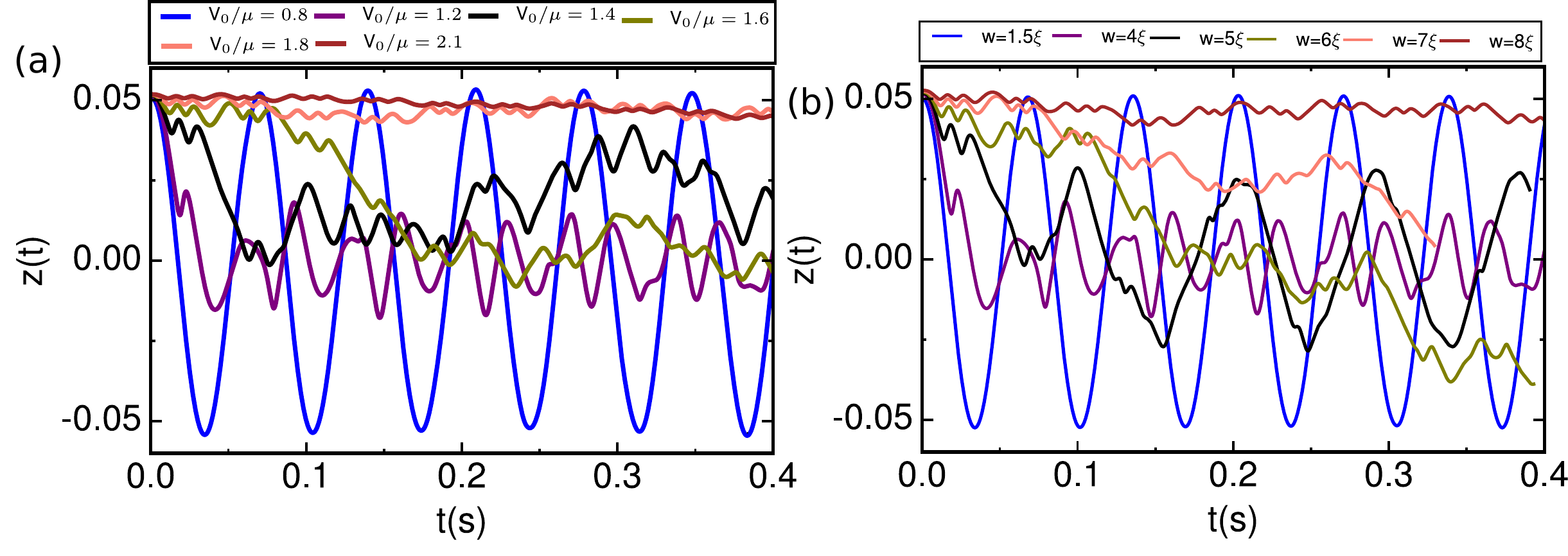}
\caption{Time evolution of the population imbalance, $z(t)$, for fixed initial imbalance $z_0$: Shown are the cases of (a) fixed $w/\xi=4$ and variable $V_0/\mu$; and (b) fixed $V_0/\mu=1.17$ and variable $w/\xi$. 
} 
\label{fig:imb_fix_z0_elong}
\end{center}
\end{figure*}

The main text has focused on the identification of the 3 key dynamical regimes of interest, namely Josephson plasma oscillations, dissipative regime, and self-trapped regime. 
As discussed, 
a convenient way to characterize such regimes is by means of the distinct dynamical population imbalance curves, $z(t)$, which reveal a plethora of relevant informations.
In this 
Appendix we discuss in more detail intricate details about the system behaviour and the transitions and crossovers between the identified regimes in 
an elongated, and an isotropic, harmonic trap.

\subsection{Elongated Trap\label{App_1a}}

Initially, we focus on the elongated trap (LENS experimental geometry 
\cite{JO5,Lidiss}). We study how $z(t)$ changes by varying the barrier parameters (height or width) at fixed initial population imbalance $z_0$ [\ref{App_1a1}], and then present further details of its behaviour in the crossover regimes [\ref{App_1a2}].

\subsubsection{
Variable Barrier Height/Width \label{App_1a1}}

\begin{figure*}[t!]
\begin{center}
\includegraphics[width= 0.98\textwidth]{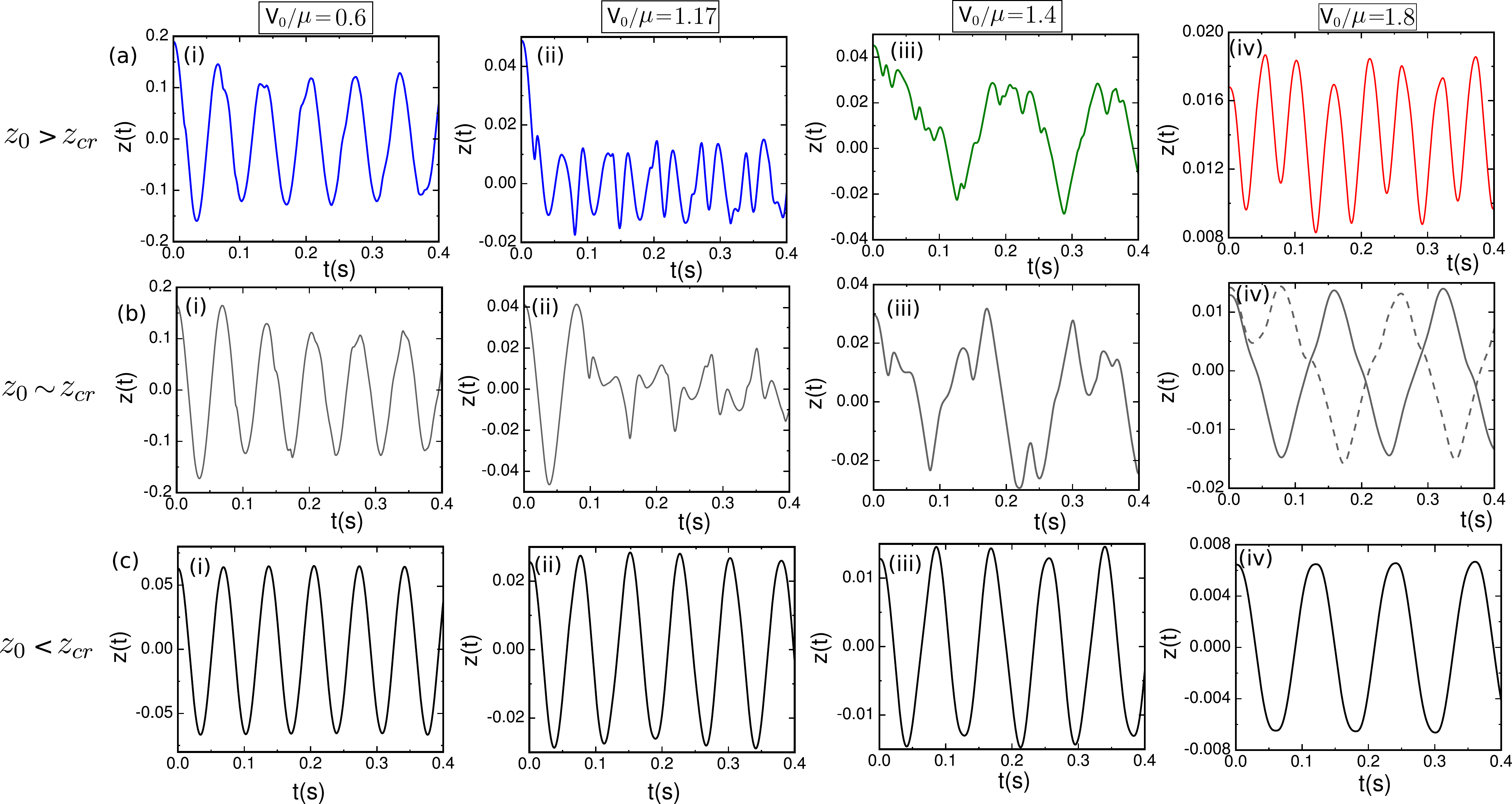}
\caption{Dependence of the temporal evolution of the population imbalance in an elongated harmonic trap on different values of $z_0/z_\mathrm{cr}$ and $V_0/\mu$ at fixed $w/\xi=4$, corresponding to the phase diagram of  Fig.~\ref{fig:w_4ksi}.
} 
\label{fig:imb_w_4ksi_elong}
\end{center}
\end{figure*}

The evolution of $z(t)$ at fixed initial population imbalance $z_0$ as the system transitions from Josephson plasma to dissipative and then to self-trapped regimes is shown in Fig.~\ref{fig:imb_w_4ksi_elong} by increasing either (a) $V_0/\mu$ at fixed $w/\xi=4$, or (b) $w/\xi$ at fixed $V_0/\mu=1.17$. This corresponds to horizontally traversing the phase diagram of, respectively, Figs.~\ref{fig:w_4ksi} and \ref{fig:zc_vs_w_v0_1_17mu}.
%
%
In both plots we see a clear transition from Josephson plasma oscillations (blue lines), to a dissipative regime (purple lines), followed by a rather complicated transition to a self-trapped state.
Due to the previously identified 
dependence of $z_{cr}$ on $V_0/\mu$ (largely related to the  profile of the maximum current versus $V_0/\mu$ found by considering first and second order terms in the tunneling Hamiltonian~\cite{Xhani20}), a simple inspection of the phase diagram of Figs.~\ref{fig:w_4ksi} and \ref{fig:zc_vs_w_v0_1_17mu} reveals that such horizontal cuts through the phase diagrams imply that the system will enter the self-trapped regime with a value of $z_0 \gg z_{cr}$, such that we do not expect to observe the emergence of the pure two-mode regime, but rather the multi-frequency self-trapped state analyzed below. 
%
Thus, these two subplots reveal the complicated intermediate region for large $z_0 > z_{cr}$ when the system transitions from the dissipative to the self-trapped regime, i.e.~the transition from the blue to the red regions in Figs.~\ref{fig:w_4ksi} and \ref{fig:zc_vs_w_v0_1_17mu} respectively.


\subsubsection{Further Dynamical Regime Details in an Elongated Trap  \label{App_1a2}}

Next, we present more details about how the evolution of $z(t)$ can differ even within a particular regime, which allows us to highlight more clearly the nature of the different observed crossovers.
In particular, Fig.~\ref{fig:imb_w_4ksi_elong} shows the complex dependence of the evolution of $z(t)$ for barriers with $w/\xi=4$ and different values of the parameters $z_0/z_{cr}$ and $V_0/\mu$, as clearly identified in the individual subplots.
Specifically, this figure is organised across 3 rows, with the ratio of $z_0/z_{cr}$ increasing from bottom to top:
from values $z_0/z_{cr}<1$ (bottom row; Josephson plasma regime), through
$z_0/z_{cr} \sim 1$ (middle row; crossover regime from Josephson to other dynamical behaviour) to $z_0/z_{cr}>1$ (top row)
but with selected values $z_0$ which are neither much smaller, nor much larger than the corresponding $z_{cr}(V_0/\mu)$.

Specifically, the bottom row (c) has  $z_0/z_{cr}<1$, with the system exhibiting Josephson plasma oscillations, independent of the value of $V_0/\mu$, but with a frequency that becomes smaller at higher $V_0/\mu$.
The intermediate row (b), depicting the behaviour of $z(t)$ at values $z_0 / z_{cr} \sim 1$, but just below 1, corresponds to the `grey' crossover regime, during which the system transitions from clean Josephson oscillations to the other identified dynamical regimes (dissipative, or self-trapped).
The top row corresponds to cases $z_0/z_{cr}>1$, for which the system exhibits a smooth crossover from dissipative regime 
(left two columns, $V_0/\mu \le 1.2$) to the self-trapped regime (right column;  $V_0/\mu =1.8$).

This figure displays a range of additional interesting behaviours, which are briefly commented upon below:

\begin{enumerate}
\item For $z_0/z_{cr}>1$ and relatively low $V_0/\mu$ [left 2 columns], the dynamical behaviour in the dissipative regime exhibits either a rapid decay of the amplitude of $z(t)$ to 0 [subplot (a)(ii)], or to a more gradual decay as $V_0/\mu$ decreases further below 1 [subplot (a)(i)]. 
\item As the system transitions from Josephson to a distinct dynamical regime with increasing $z_0$ approaching $z_{cr}$ from below -- and for intermediate values of $V_0/\mu$ of order 1 (here the value corresponds to 1.4 for the particular elongated geometry and $w/\xi =4$) -- the evolution of $z(t)$ becomes somewhat irregular, and hard to classify as belonging to a particular regime [see, e.g.~subplot (b)(iii)]. 
Such features are somewhat obscured by our chosen coloured representation of the phase diagram shown in Fig.~\ref{fig:w_4ksi} but are important to note: it is precisely such behaviour of $z(t)$ which reveals the irregular crossover denoted by different colours in the presented phase diagram plots. 
\item Another interesting feature is the 
characterization of the regime we have more broadly termed `pure self-trapped' in this work.
Pure self-trapped here refers to a well-defined regime in which there are periodic well-defined oscillations in the population imbalance with frequency $\nu _\mathrm{MQST}=\Delta \mu /h$ (where $\Delta \mu $ is the chemical potential difference between the two-condensates), whose sign however does not change, with associated running phase. 
Such self-trapping emerges in its purest form for values of $z_0$ marginally above z$_{cr}$, and for values $V_0/\mu$ sufficiently exceeding 1 for the given barrier width: the standard transition region is portrayed by the two curves in (b)(iv), which mark the onset of the transition from Josephson plasma to self-trapping~\cite{MQST1,MQST3}. 
This is discussed further in \ref{app_mqst} below.

\item In the `crossover regime' occurring for $z_0 > z_{cr}$ and `intermediate' $V_0/\mu \geq 1.2$ [subplot (a)(iii)] we see behaviour which is somewhat reminiscent of the dissipative regime, as
there are multiple dips in the early-time decay of $z(t)$, followed by a periodic-like oscillation around zero mean value; at the same time, it also bears features of the behaviour seen for high $z_0/z_{cr}$ and large $V_0/\mu >1$: indeed increasing $z_0/z_{cr}$ even more for fixed $V_0/\mu=1.4$ gradually leads to the appearance of a self-trapped state.
This behaviour further justifies the slightly curved contours shown in the phase diagrams.  
Understanding the existence of such features in the intermediate regime could potentially be useful in guiding future experimental attempts to probe the entire phase diagram.
\item Finally, we note some interesting observations in the Josephson regime $z_0<z_{cr}$ [subplots (c)(i)-(iv)].
The Josephson regime in ultracold atoms is typically associated with a single dominant frequency, the plasma frequency, determined by the two-mode model.
However, the two-mode approximation is expected to be valid if the condensate wavefunctions in the left and right wells are localized in each well, i.e.~if  the barrier height and width are such that the overlapping of the left and right condensate wavefunctions is small in the barrier region. 
As shown already in \ref{app_1} both the Josephson plasma frequency and the critical imbalance are well predicted by the two-mode model in our elongated trap, only for a specific parameter subspace.
Specifically, the predictions of the two-mode model appear to agree with those of our GPE simulations at fixed $w/\xi=4$ only for $V_0/\mu \geq 1.6$.
%
For $V_0/\mu \leq 1.4$, we actually see higher frequencies emerging, at approximate integer multiple values of the fundamental Josephson oscillation frequency, and with a significantly lower weighting, consistent with Refs.~\cite{Meier2001,Goldobin2007,zaccanti_19}.
\end{enumerate}

\begin{figure*}[t!]
\begin{center}
\includegraphics[width=0.7 \textwidth]{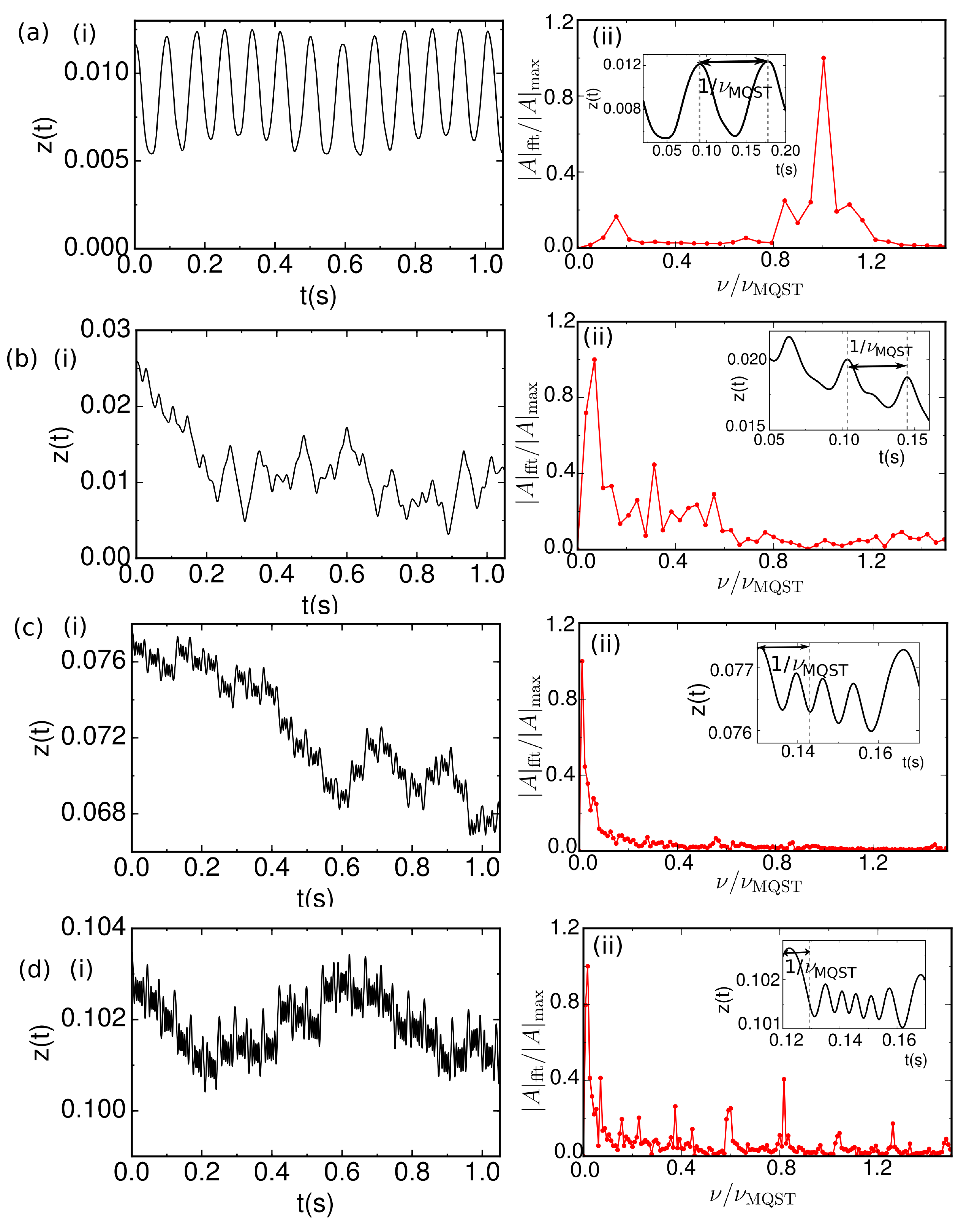}
\caption{(a)-(d)(i) Temporal evolution of the population imbalance, $z(t)$, for different  initial imbalances $z_0$ and for $V_0/\mu=2$, $w/\xi = 4$.  Plots (a)-(d)(ii) show the corresponding DFT amplitude of $z(t)-\langle z(t) \rangle$, with the frequency scaled to the MQST prediction, and the amplitude scaled to the value of the DFT amplitude of the dominant frequency. The insets show the $z(t)$ profiles of (i) for shorter time evolution where the arrows indicate the time interval $1/\nu_\mathrm{MQST}$ with $\nu _\mathrm{MQST} \simeq \Delta \mu _0 /h$ the two-mode model predicted frequency.}
\label{fig:imb_fft_v0_2mu}
\end{center}
\end{figure*}

\subsubsection{Macroscopic Quantum Self-Trapping as a Limiting Case of a More General Self-Trapped State \label{app_mqst}}

Throughout this work we have emphasized the emergence -- for relatively large $w/\xi$ and $V_0/\mu$ -- of a dynamical regime which exhibits oscillatory -- but  not necessarily periodic -- population transfer, with one well being  more populated than the other within the time interval explored.
In the limiting case of the initial population imbalance $z_0$ marginally exceeding the critical value $z_{cr}$ for the particular configuration, one recovers the well-known Macroscopic Quantum Self-Trapping which we call pure self-trapped state: this features both single-frequency biased population oscillations between the two wells, and corresponding phase slips of $2 \pi$ with the same period. As $z_0$ increases in the same system we observe a number of features, which can be clearly seen in Fig.~\ref{fig:imb_fft_v0_2mu}: %
Firstly, the population oscillations become more complicated, with the gradual emergence of numerous frequencies associated with higher-order excitations.
Correspondingly the phase slips -- which continue occurring -- are not as regular. 
Importantly, the self-trapped frequency $ \nu _\mathrm{MQST} \simeq \Delta \mu  /h$ predicted by the two-mode model becomes increasingly less relevant with higher $z_0$, with all dominant frequencies in such extended self-trapped states being slower than the corresponding two-mode model predictions.
The higher the initial $z_0$, the more reduced the total oscillation amplitude of $z(t)$ becomes. In the limit of $z_0 / z_{cr} \sim$ few, 
the evolution of $z(t)$ resembles to good approximation a nearly flat straight line with features only becoming discernible when one zooms into the plot.



\subsection{Further Characterization of Dynamical Regime Crossover for the Isotropic Trap\label{app_iso}}

\begin{figure*}[h!]
\begin{center}
\includegraphics[width= 0.98\textwidth]{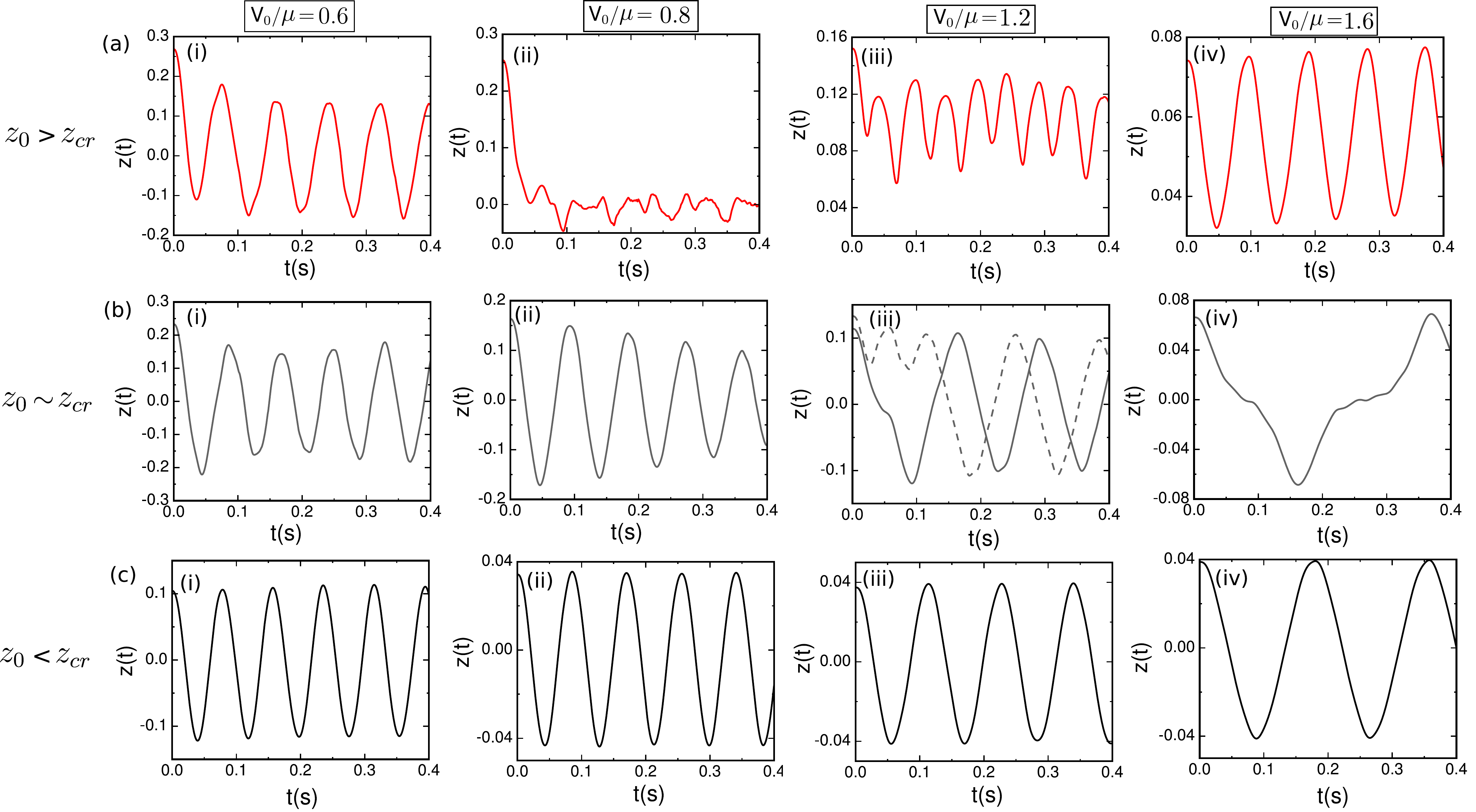}
\caption{
Dependence of the temporal evolution of the population imbalance in an isotropic trap on different values of $z_0/z_\mathrm{cr}$ and $V_0/\mu$ at fixed $w/\xi = 4$, corresponding to the phase diagram of Fig.~\ref{fig:diagram_spher}
} 
\label{fig:imb_spher}
\end{center}
\end{figure*}

The phase diagram shown in Fig.~\ref{fig:diagram_spher} for the isotropic trap revealed a broader, more pronounced crossover region (grey region) between Josephson and the other (dissipative, self-trapped) dynamical regimes.
This could be associated with the 
 enhanced coupling of the Josephson plasma oscillations to other intra-well excitations.
Fig.~\ref{fig:imb_spher} shows the complex dependence of the evolution of $z(t)$ for barriers with $w/\xi = 4$ and different values of the parameters $z_0/z_{cr}$ (increasing from bottom to top) and $V_0/\mu$ (increasing from left to right).
Overall we find similar features to the corresponding figure for the elongated trap
[Fig.~\ref{fig:imb_w_4ksi_elong}].
This demonstrates the broad applicability of our findings.


\end{document}